\documentclass[aps,prx,nobibnotes,superscriptaddress,amsfonts,amsmath,amssymb,showpacs,floatfix,reprint,longbibliography,nofootinbib]{revtex4-2}

\usepackage{natbib}
\usepackage[english]{babel}
\usepackage{letltxmacro}
\usepackage{latexsym}
\LetLtxMacro{\ORIGselectlanguage}{\selectlanguage}
\makeatletter
\DeclareRobustCommand{\selectlanguage}[1]{%
  \@ifundefined{alias@\string#1}
    {\ORIGselectlanguage{#1}}
    {\begingroup\edef\x{\endgroup
       \noexpand\ORIGselectlanguage{\@nameuse{alias@#1}}}\x}%
}
\newcommand{\definelanguagealias}[2]{%
  \@namedef{alias@#1}{#2}%
}
\makeatother
\definelanguagealias{en}{english}
\definelanguagealias{English}{english}
\usepackage{graphicx}
\usepackage{amsmath}
\usepackage{amsfonts}
\usepackage{amssymb}
\usepackage{bm}
\usepackage{color}
\usepackage[percent]{overpic}
\usepackage{soul} 
\usepackage{amssymb}
\usepackage{wasysym}
\usepackage{dsfont}
\usepackage{float}
\usepackage[mathscr]{euscript}
\usepackage{lipsum}
\usepackage{hyperref}
\hypersetup{
    pdfstartview={FitH},    
    colorlinks=true,       
    linkcolor=blue,          
    citecolor=blue,        
    filecolor=magenta,      
    urlcolor=blue           
}

\usepackage{pdfpages}
\usepackage{pgffor}
\makeatletter
\AtBeginDocument{\let\LS@rot\@undefined}
\makeatother

\newcommand{\pdagger}{\phantom{\dagger}}
\newcommand{\be}{\begin{equation}}
\newcommand{\ee}{\end{equation}}
\newcommand{\bea}{\begin{eqnarray}}
\newcommand{\eea}{\end{eqnarray}}

\newcommand{\mc}{\mathcal}

\newcommand{\vect}[1]{\mathbf{#1}}

\usepackage{graphicx}
\usepackage[colorinlistoftodos]{todonotes}
\usepackage{verbatim}
\usepackage[normalem]{ulem}
\usepackage{enumitem}
\usepackage{multirow}

\begin{document}
\title{Fermionic parton theory of Rydberg $\mathbb{Z}_2$ quantum spin liquids}

\author{Atanu Maity}
\affiliation{Institut f\"ur Theoretische Physik und Astrophysik and W\"urzburg-Dresden Cluster of Excellence ct.qmat, Julius-Maximilians-Universit\"at W\"urzburg, Am Hubland, Campus S\"ud, W\"urzburg 97074, Germany}
\affiliation{Department of Physics and Quantum Centre for Diamond and Emergent Materials (QuCenDiEM), Indian Institute of Technology Madras, Chennai 600036, India}
\author{Yasir Iqbal}
\affiliation{Department of Physics and Quantum Centre for Diamond and Emergent Materials (QuCenDiEM), Indian Institute of Technology Madras, Chennai 600036, India}
\author{Rhine Samajdar}
\email{rhine\_samajdar@princeton.edu}
\affiliation{Department of Physics, Princeton University, Princeton, NJ 08544, USA}
\affiliation{Princeton Center for Theoretical Science, Princeton University, Princeton, NJ 08544, USA}

\begin{abstract}
Programmable quantum simulators based on neutral atom arrays today offer powerful platforms for studying strongly correlated phases of quantum matter. Here, we employ the projective symmetry group framework to describe the symmetry fractionalization patterns in a topologically ordered $\mathbb{Z}_{2}$ quantum spin liquid (QSL) synthesized in such a Rydberg array on the ruby lattice. By systematically comparing the static structure factors of all possible mean-field \textit{Ans\"atze} against density-matrix renormalization group calculations, we identify a promising candidate for the precise $\mathbb{Z}_{2}$ QSL realized microscopically. We also present detailed analyses of the dynamical structure factors as a reference for future experiments and showcase how these spin correlations can differentiate between varied QSL \textit{Ans\"atze}.
\end{abstract}

\maketitle

\section{Introduction}

Quantum spin liquids (QSLs) are fascinating strongly correlated  phases of matter characterized by long-range many-body quantum entanglement~\cite{Savary-2017}. The highly entangled nature of these quantum states lies at the root of their many exotic properties, such as  emergent gauge fields~\cite{knolle2019field,broholm2020quantum}, and topological order beyond the Landau-Ginzburg paradigm of symmetry breaking~\cite{wen2017colloquium}. Another such defining feature of QSLs is the presence of fractionalized excitations carrying quantum numbers that are fractions of those borne by the original degrees of freedom of the system. 

Among the plethora of theoretically proposed QSL states, the simplest perhaps is the 
$\mathbb{Z}_2$ spin liquid~\cite{ReadSachdev91,Wen91,Sachdev92}. The $\mathbb{Z}_2$ QSL is a gapped phase of matter which preserves all space-group and time-reversal symmetries and possesses the same topological order as the celebrated toric code~\cite{kitaev2006anyons}. In recent years, much effort has been devoted towards the search for such a $\mathbb{Z}_2$ QSL in geometrically frustrated solid-state materials~\cite{lee2008end}. However, another particularly promising (and relatively new) direction is the realization of these QSL states in ``synthetic'' many-body systems. 

\begin{figure}[tb]
    \centering
    \includegraphics[width=0.8\linewidth]{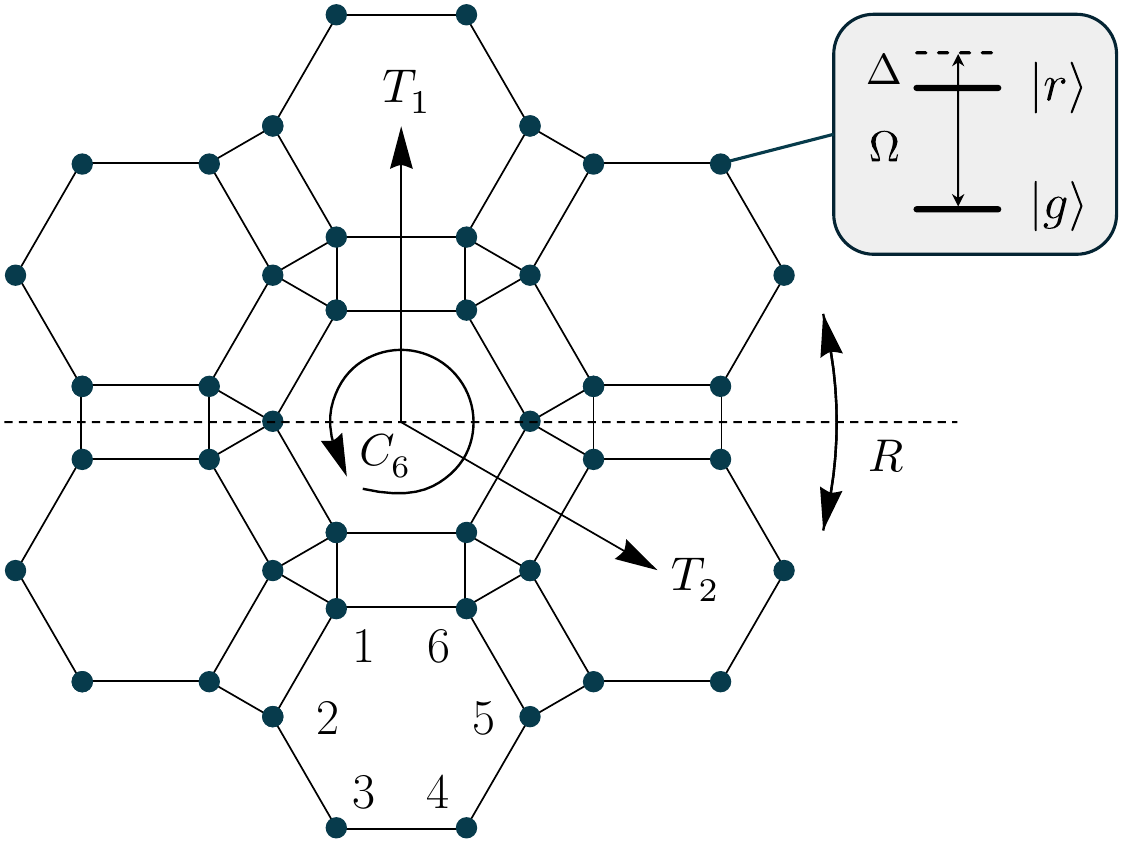}
    \caption{\textbf{Ruby lattice and the Rydberg model setup.} An illustration of the ruby lattice, highlighting the six sites in the unit cell and the space-group symmetry elements $\{T_1,T_2,C_6,R\}$. The spin at each site corresponds to a two-level system formed by the atomic ground and Rydberg states, with transitions between them driven by a laser with Rabi frequency $\Omega$ and detuning $\Delta$.}
    \label{fig:lat_structure}
\end{figure}

Strongly interacting arrays of neutral atoms offer one such versatile platform for quantum simulation of spin models~\cite{bernien2017probing}, using ensembles of optically pumped atoms interacting via excitations to high-lying Rydberg states. In these systems, a key constraint is imposed by the ``Rydberg blockade''~\cite{jaksch2000fast}, which energetically forbids the simultaneous excitation of neighboring atoms and  leads to a rich variety of physics in both one~\cite{keesling2019quantum,samajdar2018numerical,whitsitt2018quantum,PhysRevLett.122.017205,de2019observation} and two ~\cite{Samajdar_2020,Ebadi.2021,scholl2021quantum, kalinowski2021bulk, Chen_2023} dimensions. In particular, under certain conditions, blockade-allowed configurations of Rydberg atoms can be mapped~\cite{Samajdar.2021,Verresen.2020} to close-packed states of so-called quantum dimer models~\cite{RK, moessner2008quantum}. Based on this mapping, recent beautiful experiments by~\citet{Semeghini.2021} have demonstrated the synthesis of  a topologically ordered $\mathbb{Z}_2$ QSL---corresponding to a resonating valence bond state of dimers---in arrays of Rydberg atoms on the ruby lattice~\cite{Giudici-2022}.

Importantly, just the $\mathbb{Z}_2$ nature of the QSL alone does not fully specify the microscopic state; there can be many different types of $\mathbb{Z}_2$ QSLs  for the same lattice. This can be understood as a direct consequence of fractionalization: the correct degrees of freedom to describe the QSL are not the microscopic spin variables but rather the emergent fractionalized ones, termed ``partons''. 
In the parton Hilbert space, however, any physical symmetry operation can also generically be accompanied by a gauge transformation. Stated otherwise, the  symmetries act projectively in this space, and this projective action of lattice and time-reversal symmetries define and distinguish between various QSL states. 
For instance,~\citet{Wen-2002} famously showed that there are 272 distinct symmetric $\mathbb{Z}_2$ QSL states for spin-rotation-invariant square-lattice systems. Therefore, the natural question to ask is, what are the possible $\mathbb{Z}_2$ QSL states for the Rydberg array, and of these, which is the one realized microscopically?

In this work, we address this question using a two-pronged approach. The standard theoretical framework for constructing QSL states is parton mean-field theory~\cite{Wenbook,Baskaran-1987,Baskaran-1988}; adopting this approach, we describe our system in terms of emergent fermionic partons. Through a careful analysis of the projective symmetry group (PSG)~\cite{Wen-2002,wen2002quantum}, we systematically classify all allowed $\mathbb{Z}_2$ QSL states based on the symmetries of the ruby-lattice Rydberg system. A significant technical challenge in this regard (which has also previously complicated studies in strongly spin-orbit-coupled systems~\cite{Dodds-2013,Reuther-2014}) is that the Rydberg Hamiltonian (see Eq.~\eqref{eq:HRyd} below) does not possess an SU(2) spin-rotation symmetry; this introduces a large number of singlet and triplet terms to consider. Here, overcoming this difficulty, we build upon the classification scheme introduced for the spin-rotation-invariant case~\cite{msi} and investigate the effects of breaking the global SU(2) symmetry. The present analysis is thus qualitatively distinct from and significantly richer than previous ruby-lattice PSG studies restricted to Heisenberg models~\cite{msi}: allowing the longitudinal triplet channels needed to capture the physical Rydberg interactions yields classes of $\mathbb{Z}_2$ QSLs that have no counterpart in the SU(2)-symmetric case.

Based on PSG considerations, we find a total of 50 realizable $\mathbb{Z}_2$ QSLs.
Then, we perform a systematic search in the parameter space of spinon mean-field Hamiltonians to identify from among these the most likely candidate for the Rydberg QSL state. 
We further characterize the spin correlations for these states, presenting in-depth predictions for static and dynamical structure factors, which can be measured in future experiments. Our analyses not only yield valuable insights into the microscopic nature of the Rydberg $ \mathbb{Z}_2$ QSL state but are also universal in that they rely solely on symmetries rather than the precise tuning parameters of the Hamiltonian. Taken together, our results (i) provide a complete PSG classification of $\mathbb{Z}_2$ QSLs on the ruby lattice in the physically relevant broken-SU(2) setting, (ii) identify a concrete candidate microscopic state for the Rydberg ruby-lattice QSL, and (iii) furnish a comprehensive catalog of static and dynamical spin correlations against which future experiments on Rydberg arrays can be compared.

\section{Results}

\subsection{Model and parton construction}
We consider an array of $N$ Rydberg atoms positioned on the sites $\{\vect{r}_i \}$ of the ruby lattice (Fig.~\ref{fig:lat_structure}). Each atom can be regarded as a two-level spin-1/2 system where the atomic ground ($\vert g \rangle$) and Rydberg ($\vert r \rangle$) states correspond to spin down and up, respectively. Defining $S^x_i \equiv (\vert g \rangle^{}_i  \langle r \vert + \mathrm{h.c.})/2$, $n^{}_i \equiv \vert r \rangle^{}_i  \langle r \vert$, and $S^{z}_i \equiv n^{}_i - 1/2$, the  many-body Hamiltonian can be written as a long-ranged transverse-field Ising model in a longitudinal field~\cite{sachdev2002mott,fendley2004competing}:
  \begin{align}
  \label{eq:HRyd}
    H&=\Omega\sum_{i}S^x_i -\Delta\sum_{i} n^{}_i+\sum_{\langle i,j\rangle}V^{}_{ij} n^{}_i n^{}_j.
\end{align}
Here, $\Omega$ is the Rabi frequency of the laser that drives transitions between the ground and Rydberg states, $\Delta$ is the laser detuning, and $V_{ij} = V_0 /\vert \vert \vect{r}_i - \vect{r}_j \vert \vert^6$ is a long-ranged van der Waals potential arising from the strong dipole-dipole interactions~\cite{Lukin.2000, bernien2017probing}.

On choosing the interatomic spacing such that up to third-nearest-neighboring atoms are mutually blockaded, the Rydberg Hamiltonian on the ruby lattice is known to host a $\mathbb{Z}_2$ quantum spin liquid ground state~\cite{Semeghini.2021}. The identification of this state as a QSL in the experiment of Ref.~\cite{Semeghini.2021} is in fact subtle: truly infinite-range van der Waals tails can destabilize the QSL in favor of a direct transition from the paramagnet to a valence bond solid~\cite{Verresen.2020}, but in the nonequilibrium setting of the experiment, the system's metastable phenomenology remains consistent with the QSL phase of the short-range-truncated model that we consider here~\cite{Giudici-2022}. This $\mathbb{Z}_2$ QSL corresponds to the deconfined phase of a gauge theory~\cite{Samajdar-2023, tarabunga2022gauge} and supports three species of quasiparticle excitations~\cite{read1989statistics}: a bosonic ``spinon'' (called the $e$ particle in the language of the toric code), a bosonic ``vison'' ($m$)~\cite{SenthilFisher}, and a fermionic spinon ($\varepsilon = e \times m$). A full classification of the symmetry-enriched $\mathbb{Z}_2$ topological order requires tracking the symmetry fractionalization on all of these anyonic sectors~\cite{essin2013classifying,Barkeshli-2019,Ye-Zou-2024}; in the Rydberg system, the three sectors reside on three different lattices (kagome for $e$, dice for $m$, and ruby for $\varepsilon$), and so each would need to be addressed by a separate PSG analysis on the appropriate lattice. The fermionic parton construction that we employ here targets the $\varepsilon$ sector on the ruby lattice; we comment on the bosonic counterparts~\cite{Yang-2016} in Sec.~\ref{sec:discussion}. The time-honored method which facilitates the study of such fractionalized excitations and other emergent properties of the QSL state is the parton construction.

In the parton framework, the physical spins are decomposed into more fundamental degrees of freedom, which carry spin but no charge, thus allowing for a natural description of the fractionalization of quantum numbers. Mathematically, the spin operators on each lattice site are expressed in terms of two flavors of complex spin-$1/2$ fermionic quasiparticles (known as Abrikosov fermions). Denoting the two types of partons using a pseudospin index $\sigma$\,$=$\,${\uparrow, \downarrow}$, we can write 
$
{S}^{\gamma}_{i}$\,$=$\,$({f}^\dagger_{i\sigma}\tau^{\gamma}_{\sigma\sigma^\prime}{f}^{\pdagger}_{i\sigma^\prime})/2,
$
where $\tau^\gamma$ ($\gamma=x,y,z$) is a Pauli matrix, and summation is implied over repeated indices. Such a rewriting enlarges the local Hilbert space, so, to remain in the physical subspace, we impose an additional  constraint 
${f}_{i\sigma}^{\dagger}{f}^{\pdagger}_{i\sigma} = 1, \;  {f}^{\pdagger}_{i\sigma}{f}^{\pdagger}_{i\sigma'}\epsilon^{\pdagger}_{\sigma\sigma'} = 0
$~\cite{Baskaran-1987,Baskaran-1988},
ensuring that there is exactly one fermion on each site.
Conveniently, the spin operator can now be recast using a spinor doublet ${\psi}_i=({\phi}_i,{\bar{\phi}}_i)$, where ${\phi}_i$\,$=$\,$({f}_{i,\uparrow},{f}_{i,\downarrow})^\textsc{T}$ and ${\bar{\phi}}_i$\,$=$\,$({f}^\dagger_{i,\downarrow},-{f}^\dagger_{i,\uparrow})^\textsc{T}$, as 
$
{S}^{\gamma}_{i}=\text{Tr}[{\psi}^\dagger_i\tau^\gamma{\psi}^{\pdagger}_i]/2
$~\cite{Affleck-1988b};
the associated half-filling constraint  then reads
${\psi}^{\pdagger}_{i}\tau^{\gamma}{\psi}^{\dagger}_{i}$\,$=$\,$0.
$
A key consequence of this formulation is the emergence of a local gauge symmetry absent in the original spin Hamiltonian: the spin operators remain invariant under a site-dependent transformation ${\psi}_i \rightarrow {\psi}_i W^{}_i$, for $W^{}_i \in \text{SU(2)}$. As we show below, this emergent gauge symmetry bears important implications for the action of physical symmetry operations.

\begin{table*}    
\begin{ruledtabular}
\begin{tabular}{cccccccccccc}
\multirow{2}{*}{No.}& \multirow{2}{*}{$\{\eta,\eta_{C_6},\eta_R,\eta_{C_6R},\eta_{\Theta C_6}\}$}& \multirow{2}{*}{$g^{}_\Theta$} & \multicolumn{2}{c}{1NN} & \multicolumn{2}{c}{2NN} & %
    \multicolumn{2}{c}{3NN} & \multicolumn{2}{c}{Onsite} & \multirow{2}{*}{IGG}  \\
\cline{4-5}
\cline{6-7}
\cline{8-9}
\cline{10-11}
& & & $\mc{U}^{0}_{1}$ & $\mc{U}^{z}_{1}$ & $\mc{U}^{0}_{2}$ & $\mc{U}^{z}_{2}$ & $\mc{U}^{0}_{3}$ & $\mc{U}^{z}_{3}$ & $\mc{U}^{0}_{0}$ &$\mc{U}^{z}_{0}$  \\
\hline
1& $\{\eta,\pm \eta,+,+,+\}$ & $\dot\iota\tau^y$ & $\tau^{x,z}$ & $\tau^0$ & $\tau^{x,z}$ & $\tau^0$ & $\tau^{x,z}/0$ & $\tau^0/\dot\iota\tau^y$ & $\tau^{x,z}$ & $\tau^0$ & $\mathds{Z}_2$ \\
2& $\{\eta,\pm \eta,+,-,+\}$ & $\dot\iota\tau^y$ & $\tau^{x,z}$ & $\tau^0$ & $0$ & $\dot\iota\tau^y$ & $\tau^{x,z}/0$ & $\tau^0/\dot\iota\tau^y$ & $\tau^{x,z}$ & $\tau^0$ & $\mathds{Z}_2$ \\
3&$\{\eta,\pm \eta,+,+,-\}$ & $\dot\iota\tau^y$ & $\tau^{x,z}$ & $\tau^0$ & $\tau^{y}$ & $0$ & $\tau^{y}/\dot\iota\tau^0$ & $0/\dot\iota\tau^{z,x}$ & $\tau^{x,z}$ & $\tau^0$ & $\mathds{Z}_2$ \\
4&$\{\eta,\pm \eta,+,-,-\}$ & $\dot\iota\tau^y$ & $\tau^{x,z}$ & $\tau^0$ & $\dot\iota\tau^{0}$ & $\dot\iota\tau^{x,z}$ & $\tau^{y}/\dot\iota\tau^0$ & $0/\dot\iota\tau^{z,x}$ & $\tau^{x,z}$ & $\tau^0$ & $\mathds{Z}_2$ \\
5& $\{\eta,\pm \eta,-,+,+\}$ & $\dot\iota\tau^y$ & $\tau^{z}$ & $\tau^0,\dot\iota\tau^{y}$ & $\tau^{z}$ & $\tau^0,\dot\iota\tau^{y}$ & $\tau^{x,z}/0$ & $\tau^0/\dot\iota\tau^y$ & $\tau^{z}$ & $\tau^0$ & $\mathds{Z}_2$ \\
6& $\{\eta,\pm \eta,-,-,+\}$ & $\dot\iota\tau^y$ & $\tau^{z}$ & $\tau^0,\dot\iota\tau^{y}$ & $\tau^{x}$ & $0$ & $\tau^{x,z}/0$ & $\tau^0/\dot\iota\tau^y$ & $\tau^{z}$ & $\tau^0$ & $\mathds{Z}_2$ \\
7& $\{\eta,\pm \eta,-,+,-\}$ & $\dot\iota\tau^y$ & $\tau^{z}$ & $\tau^0,\dot\iota\tau^{y}$ & $0$ & $\dot\iota\tau^{x}$ & $\tau^{y}/\dot\iota\tau^{0}$ & $0/\dot\iota\tau^{x,z}$ & $\tau^{z}$ & $\tau^0$ & $\mathds{Z}_2$ \\
8& $\{\eta,\pm \eta,-,-,-\}$ & $\dot\iota\tau^y$ & $\tau^{z}$ & $\tau^0,\dot\iota\tau^{y}$ & $\dot\iota\tau^0,\tau^{y}$ & $\dot\iota\tau^{z}$ & $\tau^{y}/\dot\iota\tau^{0}$ & $0/\dot\iota\tau^{x,z}$ & $\tau^{z}$ & $\tau^0$ & $\mathds{Z}_2$ \\
9&$\{\eta,\pm \eta,-,-,+\}$ & $\dot\iota\tau^z$ & $0$ & $\tau^0$ & $\tau^{x,y}$ & $\dot\iota\tau^{z}$ & $\tau^{x,y}/0$ & $\tau^0/\dot\iota\tau^z$ & $0$ & $\tau^0$ & $\mathds{Z}_2$ \\
10& $\{\eta,\pm \eta,-,+,-\}$ & $\dot\iota\tau^z$ & $0$ & $\tau^0$ & $\tau^{z}$ & $\dot\iota\tau^{x,y}$  & $\tau^{z}/\dot\iota\tau^0$ & $0/\dot\iota\tau^{x,y}$ & $0$ & $\tau^0$ & $\mathds{Z}_2$ \\
11& $\{\eta,\pm \eta,-,+,-\}$ & $\tau^0$ & $0$ & $\tau^0,\dot\iota\tau^{x,y}$ & $\tau^{z}$ & $0$ & $\tau^{x,y,z}/\dot\iota\tau^0$ & $0$ & $0$ & $\tau^0$ & $\mathds{Z}_2$ \\
12& $\{\eta,\pm \eta,-,-,-\}$ & $\tau^0$ & $0$ & $\tau^0,\dot\iota\tau^{x,y}$ & $\dot\iota\tau^{0},\tau^{x,y}$ & $0$ & $\tau^{x,y,z}/\dot\iota\tau^0$ & $0$ & $0$ & $\tau^0$ & $\mathds{Z}_2$  \\
 \hline
13& $\{\eta,\pm \eta,+,+,-\}$ & $\tau^0$ & $0$ & $\tau^0$ & $\tau^{x,y,z}$ & $0$ & $\tau^{x,y,z}/\dot\iota\tau^0$ & $0$ & $0$ & $\tau^0$ & $\mathds{Z}_2/$U(1)\\
14& $\{\eta,\pm \eta,+,-,-\}$ & $\tau^0$ & $0$ & $\tau^0$ & $\dot\iota\tau^0$ & $0$ & $\tau^{x,y,z}/\dot\iota\tau^0$ & $0$ & $0$ & $\tau^0$ & U(1) \\
15& $\{\eta,\pm \eta,-,+,+\}$ & $\dot\iota\tau^z$ & $0$ & $\tau^0$ & $0$ & $\tau^0$ & $\tau^{x,y}/0$ & $\tau^0/\dot\iota\tau^z$ & $0$ & $\tau^0$ & U(1)  \\
16& $\{\eta,\pm \eta,-,-,-\}$ & $\dot\iota\tau^z$ & $0$ & $\tau^0$ & $\dot\iota\tau^{0}$ & $0$ & $\tau^{z}/\dot\iota\tau^0$ & $0/\dot\iota\tau^{x,y}$ & $0$ & $\tau^0$ & U(1)\\
\end{tabular}
\end{ruledtabular}
\caption{\label{table:ansatze}All possible $\{T_1,T_2,R,C_6,\Theta\}\times  \mathrm{U}(1)_{\text{spin}}$-symmetric mean field \textit{Ans\"atze} with up to third-nearest-neighbor hopping and pairing terms. In the chosen gauge, the onsite terms are uniform, i.e., $u^{\alpha}_{ii}=\mc{U}^{\alpha}_{0}$. Each row gives rise to four \textit{Ans\"atze} with different combinations of $\eta=\pm1$, $\eta_{C_6}=\pm\eta$. In the ``3NN'' and ``IGG'' columns above, the term on the left (right) side of ``/" corresponds to $\eta_{C_6}= + \eta$ ($\eta_{C_6}= - \eta$).}
\end{table*}

\subsection{Mean-field \textit{Ans\"atze}}
Next, we insert this fermionic representation of the spin operators into the Hamiltonian~\eqref{eq:HRyd}. Any two-spin interaction term (e.g., $n_i n_j$) in the original model, however, becomes a four-fermion term in the parton representation, so to make further analytical progress, we perform a quadratic mean-field decomposition~\cite{Wen-2002}. There are two types of (matrix-valued) mean fields: singlets, $u^{0}_{ij}=\langle{\psi}^\dagger_i{\psi^{\pdagger}_j}\rangle$, and (three) triplets, $u^{\gamma}_{ij}=\langle{\psi}^\dagger_i\tau^\gamma{\psi^{\pdagger}_j}\rangle$. In terms of these fields, any generic spin Hamiltonian can be written in the quadratic form~\cite{Liu-2021} (see Supplementary Information  (SI)~\cite{sm} for details): 
\begin{align}\label{eq:mf_ham_singlet_triplet}
H_{q} = & \sum_{\langle i,j \rangle} \text{Tr}\left[\tau^\alpha{\psi}^{\pdagger}_iu^{\alpha}_{ij}{\psi}^\dagger_j + \mathrm{h.c.} \right] + \sum_{i} \text{Tr}\left[\tau^\alpha{\psi}^{\pdagger}_{i} u^{\alpha}_{ii}{\psi}^{\dagger}_{i}\right ],\;
\end{align}
where $\alpha$\,$=$\,$0$ ($\alpha$\,$=$\,$x,y,z$) corresponds to singlet (triplet) fields. The set of expectation values $\{u^{\alpha}_{ij}\}$ define an  \textit{Ansatz} for QSL states. Each $u^{\alpha}_{ij}$ can be parametrized using a basis of  Pauli matrices as
$ u^{0}_{ij} =  \dot{\iota}s^0_{ij} \tau^0+s^1_{ij} \tau^x +s^2_{ij}\tau^y  + s^3_{ij} \tau^z,$ and $
u^{\gamma}_{ij} =  t^{\gamma,0}_{ij}\tau^0+\dot{\iota}(t^{\gamma,1}_{ij}\tau^x+t^{\gamma,2}_{ij}\tau^y+t^{\gamma,3}_{ij}\tau^z ), \;$
for real coefficients $s_{ij}$, $t_{ij}$. In this basis, satisfying the requirement that $u^{\alpha}_{ji}=u^{\alpha}_{ij}{}^\dagger$ restricts certain onsite terms as $s^0_{ii}$\,$=$\,$t^{\gamma,1}_{ii}$\,$=$\,$t^{\gamma,2}_{ii}$\,$=$\,$t^{\gamma,3}_{ii}$\,$=$\,$0$. Note that $u^{0}_{ii}$ incorporates the constraint of having one fermion per site at a mean-field level, and  $s^{1}_{ii}$, $s^{2}_{ii}$, $s^{3}_{ii}$ are the associated Lagrange multipliers.

So far, while we have written down the most generic possible quadratic parton Hamiltonian in Eq.~\eqref{eq:mf_ham_singlet_triplet}, we are yet to specify the spatial structure of $u^{\alpha}_{ij}$. In order to do so, we need to consider the symmetries of the system, which will significantly constrain the terms allowed.

\subsection{PSG classification} 

 The space group of the ruby lattice is  $p6m$ and is generated by two lattice translations ($T_1,T_2$), a sixfold rotation ($C_6$) around an axis perpendicular to the plane, and a reflection ($R$) about the $x$-axis (Fig.~\ref{fig:lat_structure}). On the Hamiltonian, these lattice symmetries act as ${S}^\gamma_i\rightarrow{S}^\gamma_{\mathcal{O}(i)}$ with $\mathcal{O}$\,$\in$\,$\{T_1,T_2,C_6,R\}$. Note that since the ``spin'' here simply represents a binary degree of freedom encoded in two atomic levels, it does not transform as a pseudovector.
Furthermore, one can identify an anti-unitary time-reversal symmetry $\Theta$ under which ${S}^\gamma_i$\,$\rightarrow$\,$(-)^{\delta_{\gamma,y}}{S}^\gamma_{i}$; this may be understood as the combination of ``conventional'' time reversal (which sends $\textbf{S}_i $\,$\rightarrow$\,$- \textbf{S}_i$) and a global spin rotation by $\pi$ about the $S^y$-axis.

Acting with the operator $ {S}^x_i$ on the Rydberg QSL creates two $e$ particles~\cite{Samajdar-2023}, resulting in an excited state which is orthogonal to the ground state. Hence, we can approximate $\langle {S}^x_i\rangle$\,$\approx$\,$0$ (although $\langle \prod_{i \in \ell }S^x_i\rangle$ along a closed loop $\ell$ may be nonzero). Consequently, all terms in the Hamiltonian~\eqref{eq:mf_ham_singlet_triplet} which would contribute to a nonvanishing expectation value of $S^x$ can be reasonably neglected. These mean fields are precisely the  transverse triplet components: $u^{x}_{ij}$ and $u^{y}_{ij}$. This allows us to consider \textit{Ans\"atze} consisting of only singlet ($u^{0}_{ij}$) and longitudinal triplet ($u^{z}_{ij}$) fields; the latter lead to $\langle {S}^z_i\rangle \ne 0$, as required for a  nonzero Rydberg excitation density. Thus, while the lattice Hamiltonian does not possess an SU(2) spin-rotation symmetry, at the level of the fermionic mean-field wavefunction, the restriction of the triplet fields to only longitudinal components yields an emergent $\text{U(1)}_\text{spin}$ symmetry.

Having established the symmetry group, our next step is to identify all possible \textit{Ans\"atze} that are consistent with this set of symmetries. However, the definition of symmetry invariance in the parton Hilbert space is rather nontrivial due to  the aforementioned SU(2) gauge freedom. To wit, under the transformation ${\psi}_i\rightarrow{\psi}_iW_i$ and $u^{\alpha}_{ij}\rightarrow W^\dagger_iu^{\alpha}_{ij}W^{\pdagger}_j$, with $W^{}_i \in \text{SU(2)}$, ${H}_{q}$ remains invariant, so two \textit{Ans\"atze} $u^{\alpha}_{ij}$ and $\Tilde{u}^{\alpha}_{ij}=W^\dagger_iu^{\alpha}_{ij}W^{\pdagger}_j$ represent the same physical state. This property defines the projective action of symmetry operations as follows. Consider an element of the symmetry group $\mathcal{O}$ which acts on an \textit{Ansatz}  as $\mathcal{O}:u^{\alpha}_{ij}$\,$\rightarrow$\,$\Tilde{u}^{\alpha}_{ij}$. 
Even if $u^{\alpha}_{ij}\neq\Tilde{u}^{\alpha}_{ij}$, the symmetry under $\mathcal{O}$ is preserved if $\exists\, W_{\mathcal{O},i}\in$ SU(2) such that $u^{\alpha}_{ij}$\,$=$\,$W_{\mathcal{O},i}^\dagger\Tilde{u}^{\alpha}_{ij}W^{\pdagger}_{\mathcal{O},j}$. 
The combined operation of $\mathcal{O}$ and its accompanying SU(2) gauge transformation $W_{\mathcal{O},i}$ defines the PSG.
Therefore, the way to identify different \textit{Ans\"atze} representing distinct QSL states at the mean-field level is to find all possible gauge-inequivalent  projective symmetry actions $W_{\mathcal{O},i}$ for $\mathcal{O}\in\{\mathds{1},T_1,T_2,C_6,R,\Theta\}$ ($\mathds{1}$ being the identity element). The projective action of the identity $W_{\mathds{1}}$ defines a subgroup of the PSG known as the invariant gauge group (IGG), i.e., the group of operations that leave the \textit{Ansatz} unchanged. The IGG determines the nature of low-energy fluctuations about the mean-field solutions, and---depending on the allowed mean-field parameters---is broken down here from SU(2) to either  U(1) or $\mathds{Z}_2$.

Expanding on a recent PSG classification~\cite{msi} of spin-liquid states on the ruby lattice, here, we determine the possible symmetric mean-field \textit{Ans\"atze} for the Rydberg QSL. We note that the definition of time reversal in Ref.~\onlinecite{msi} as $\mathcal{T}:\hat{S}^\gamma_i$\,$\rightarrow$\,$-\hat{S}^\gamma_{i}$ differs from $\Theta$, but since the lattice coordinates do not change under either $\mc{T}$ or $\Theta$ and all other symmetry operators commute with both of them, we can use the same PSG classification. Whereas Ref.~\onlinecite{msi} assumed the full space-group $\times$ global SU(2) symmetry of a Heisenberg model and therefore retained only singlet mean fields, the physical symmetry of the Rydberg system is instead the space group combined with the emergent $\mathrm{U}(1)_{\text{spin}}$ symmetry identified above. 

Table~\ref{table:ansatze} lists all such \textit{Ans\"atze} for up to third-nearest-neighbor (3NN) couplings including, crucially, the triplet terms absent for conventional Heisenberg models~\cite{msi}. The inclusion of such channels that are forbidden in the SU(2)-symmetric setting allows for classes of $\mathbb{Z}_2$ QSLs with no analogue in Ref.~\onlinecite{msi}.  The choice to truncate the PSG analysis at 3NN is motivated by two physical considerations: the strong $1/r^{6}$ decay of the van der Waals interaction, which suppresses the amplitudes of longer-range mean fields, and the fact that 3NN spans the Rydberg-blockade range directly responsible for stabilizing the QSL phase. The PSG classification itself is determined solely by symmetry and is therefore insensitive to the microscopic range of $V_{ij}$; extending the analysis to 2NN or to 4NN and beyond would only restrict or enlarge, respectively, the set of symmetry-allowed amplitudes within each class, without introducing new symmetry-distinct PSG classes. 
Every \textit{Ansatz} is labeled by a set of parameters $\{\eta,\eta_{C_6},\eta_R,\eta_{C_6R},\eta_{\Theta C_6}\}$, each of which can take values $\pm 1$, with $\eta_\mc{O}$ originating from the gauge-enriched symmetry relations of the operation $\mc{O}$~\cite{sm}. Additionally, one also has to  specify a Pauli matrix associated with the action of time-reversal symmetry, which we denote by $g_\Theta$ in Table~\ref{table:ansatze}.

Although we begin with 64 $\mathds{Z}_2$ projective extensions of the symmetry group, we find that only 50 of them can be realized with a $\mathds{Z}_2$ IGG---which is the case of interest given the $\mathbb{Z}_2$ nature of the Rydberg QSL;  the remaining correspond to a U(1) IGG. As Table~\ref{table:ansatze} conveys, the whole \textit{Ansatz} can be constructed from the knowledge of a few reference mean fields. Let us define the reference onsite, 1NN, 2NN, and 3NN fields as $\mc{U}^\alpha_{0}$, $\mc{U}^\alpha_{1}$, $\mc{U}^\alpha_{2}$, and $\mc{U}^\alpha_{3}$, respectively (see SI~\cite{sm}); these can be read off from Table~\ref{table:ansatze}. Then, $u^\alpha_{ij}$ on all other bonds of the lattice can be obtained from these eight simply by symmetry, as detailed in the SI~\cite{sm}.

\begin{figure}[tb]
    \centering
    \includegraphics[width=\linewidth]{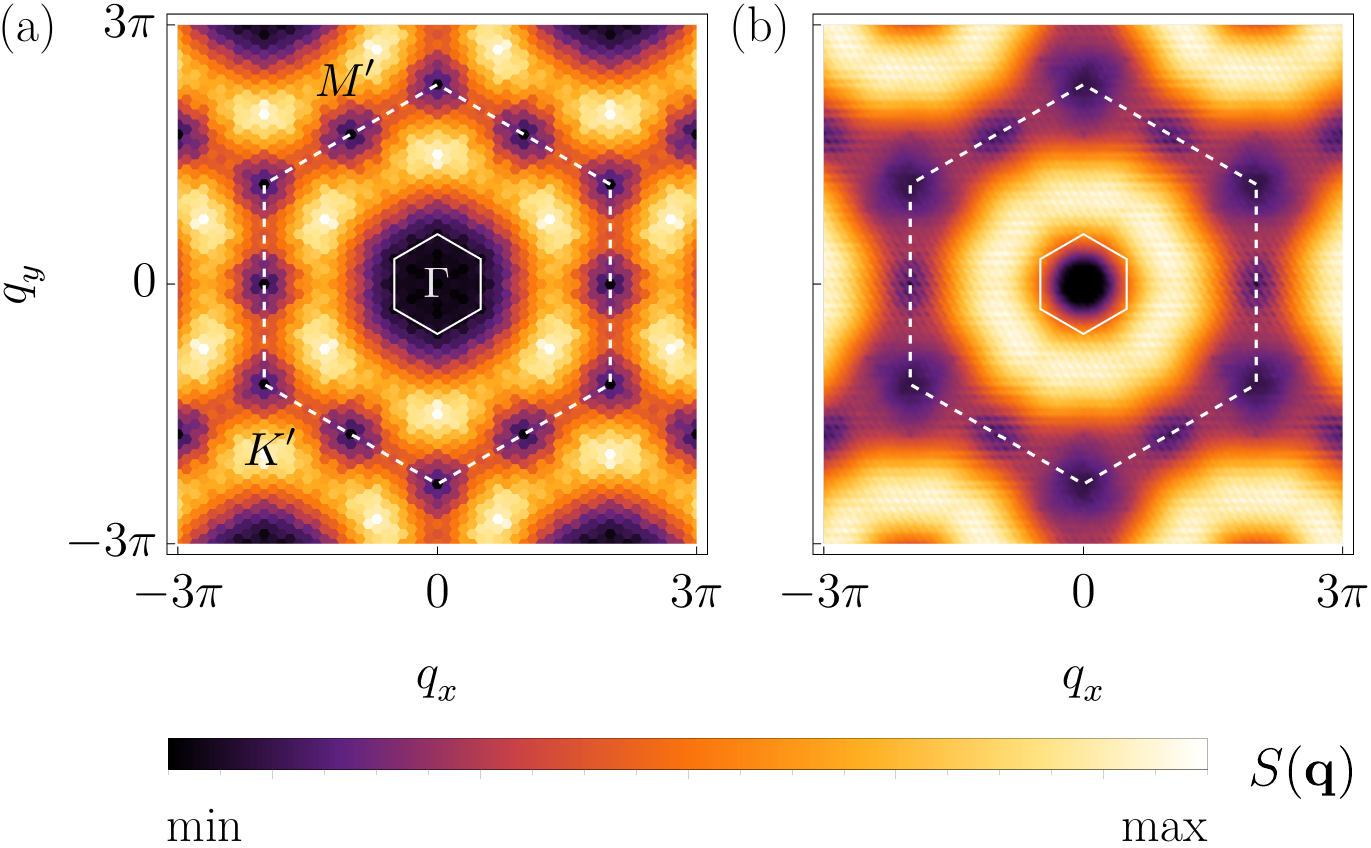}
    \caption{\textbf{Static structure factor of the candidate Rydberg $\mathbb{Z}_2$ quantum spin liquid.} (a) Static structure factor (SSF) of the Rydberg $\mathbb{Z}_2$ QSL, as determined from the DMRG ground state on a 288-site lattice. (b) SSF for the \textit{Ansatz} in row 8 of Table~\ref{table:ansatze} with $\eta = -\eta^{}_{C_6} = 1$ for (optimized) mean-field amplitudes chosen to be $|\mc{U}^{\alpha}_2|/|\mc{U}^{\alpha}_1| = r = 1.25$, $t^{z,\alpha}_{ij} = 2 s^{0,\alpha}_{ij}$, and $t^{z,0}_{ii} = 0.35$. While there are quantitative differences between the two SSFs, the qualitative similarity in the locations of their minima and maxima is apparent. The dashed (solid) hexagons mark the extended (first) Brillouin zone.}
    \label{fig:SF}
\end{figure}

\subsection{Spin correlations of the Rydberg spin liquid}
With the full list of possible QSL states in hand, we now turn to examining the experimental signatures of the different $\mathbb{Z}_2$ \textit{Ans\"atze}. 
To this end, we compute the static structure factor (SSF) $S(\mathbf{q})$, which is the Fourier transform of the equal-time spin-spin correlation function; here, we specifically look at the $z$-component $\langle S^z_i S^z_j\rangle$ since measurements in the computational basis are the most convenient experimentally.

To begin, we evaluate $S(\mathbf{q})$ for the ground state of the Rydberg Hamiltonian in the QSL phase, as obtained on a 288-site lattice using the density-matrix renormalization group (DMRG) algorithm; full details of the simulations are provided in the Methods. We work at $\Delta/\Omega = 1.7$ and $V_{ij}/\Omega = 50$ for all pairs with separation up to the third-nearest neighbor (with $V_{ij}=0$ beyond this range), which places the system well inside the $\mathbb{Z}_2$ QSL phase identified in previous studies of this effective model~\cite{Verresen.2020,Giudici-2022,chen2025anomalous}. These values are representative of the regime relevant to current Rydberg-array experiments, where $\Omega/(2\pi)$ is typically of order a few MHz, $\Delta/\Omega \sim \mathcal{O}(1)$, and the nearest-neighbor interaction $V_0/\Omega$ reaches values of several tens at the relevant atomic spacings; our aim is to characterize the universal properties of the QSL phase over this extended parameter range rather than to fit any single experimental realization. The $\mathbb{Z}_2$ topological order of the resulting ground state has been corroborated independently: the topological entanglement entropy saturates to $\gamma\simeq \ln 2$ over a broad window of $\Delta/\Omega$~\cite{Semeghini.2021,Verresen.2020,chen2025anomalous}. The SSF, shown in Fig.~\ref{fig:SF}(a), does not display any sharp Bragg peaks, as expected for a QSL, but does reveal two prominent features. First, the local minima of $S(\mathbf{q})$ are located not only at the $\Gamma\,$(center) and $K'\,$(corner) points of the extended Brillouin zone, but also at the $M'\,$(edge) points. Secondly, the SSF exhibits a local maximum along the lines connecting the $\Gamma$ and $K'$ points.

These observations suggest a route towards identifying which \textit{Ansatz} likely describes the microscopic QSL state realized for the Rydberg Hamiltonian. To do so, we focus on mean-field parameters which are set to a uniform value on all the 1NN bonds, and likewise for the 2NN bonds, but with the ratio between the two chosen to be $1:r$. Systematically tuning $r$, we then compute $S(\mathbf{q})$ for each \textit{Ansatz} and inspect it for the properties identified above. Examining the nearly 300 SSFs in the SI, we find one promising QSL state (labeled $8\{+,-,-,-,-\}$ in the notation of Table~\ref{table:ansatze}) that is consistent with those features in a certain parameter regime.  Optimizing further over the ratio of the singlet to triplet strengths for this candidate state, we see that  it indeed shows good agreement with the characteristic features of the numerically computed SSF, as presented in Fig.~\ref{fig:SF}(b).

\begin{figure}[tb]
    \centering
    \includegraphics[width=\linewidth]{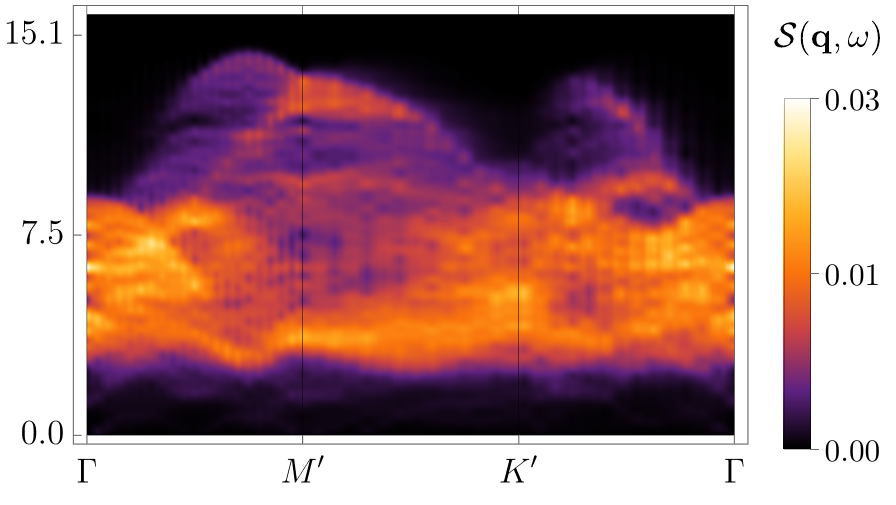}
    \caption{\textbf{Dynamical structure factor of the candidate Rydberg $\mathbb{Z}_2$ quantum spin liquid.} Dynamical structure factor $\mathcal{S}(\mathbf{q},\omega)$, plotted along a high-symmetry path in the extended Brillouin zone, for the \textit{Ansatz} $8\{+,-,-,-,-\}$ with the same parameters as in Fig.~\ref{fig:SF}(b). The diffuse continuum and the absence of sharp dispersive features are characteristic of fractionalized excitations in the $\mathbb{Z}_2$ QSL phase.}
    \label{fig:DSF}
\end{figure}

Finally, we also study the dynamical structure factor (DSF):
$
\mathcal{S}(\mathbf{q},\omega)\equiv \int^{+\infty}_{-\infty}dt/(2\pi N^2)\, e^{\dot{\iota}\omega t} \sum_{i,j}e^{\dot{\iota}\mathbf{q}\cdot(\mathbf{r}_{i}-\mathbf{r}_{j})}\times$ $[\langle {S}^z_{i}(t){S}^{z}_{j}(0)\rangle- \langle {S}^z_{i}(t)\rangle \langle{S}^{z}_{j}(0)\rangle]$.
The $\omega$-integrated version of this quantity is simply the SSF, but $\mathcal{S}(\mathbf{q},\omega)$ allows us to probe $\omega$-resolved information about the excitation spectrum characterizing the low-energy response of the QSL state.
Figure~\ref{fig:DSF} illustrates the DSF for the candidate $8\{+,-,-,-,-\}$ state (the DSFs of all other $\mathbb{Z}_2$ \textit{Ans\"atze} are tabulated in the SI for various values of $r$~\cite{sm}). Clearly, we see that the state is gapped and that the scattering is diffuse with no sharp dispersive features. 
The fact that the DSF exhibits such a broad continuum can be interpreted as a distinctive signature of fractionalization~\cite{PhysRevB.68.134424}.

\section{Discussion} 
\label{sec:discussion}
Motivated by the observation of a $\mathbb{Z}_2$ QSL in recent experiments on Rydberg atom arrays, in this work, we present a detailed projective symmetry group analysis of all possible mean-field \textit{Ans\"atze} on the ruby lattice that could describe this state. We characterize these \textit{Ans\"atze} by computing their associated static and dynamical structure factors and show how the low-energy spin correlations can distinguish between different  QSL states. Our work should thus serve as a  guide to future experiments, in which one could, say, measure the DSF~\cite{uhrich2017noninvasive} and compare it against our comprehensive listing in the SI~\cite{sm} to identify the microscopic QSL state. In fact, here, we identify a promising candidate state using a similar procedure but based on numerical (DMRG) rather than experimental input. For comparison, it would also be interesting to theoretically calculate the structure factors of the Rydberg QSL using large-scale quantum Monte Carlo simulations~\cite{yan2022triangular,wang2024renormalizedclassicalspinliquid} of kagome-lattice dimer models~\cite{misguich2002quantum,hwang2024vison}.

Beyond structure factors, a number of complementary experimental probes can sharpen the identification of the QSL phase and help discriminate between different \textit{Ans\"atze}. Nonlocal string- and loop-order operators, such as those employed in Ref.~\cite{Semeghini.2021} to diagnose $\mathbb{Z}_2$ topological order, can be measured projectively in Rydberg arrays and are particularly natural in our setting. Dynamical information, analogous to that encoded in $\mathcal{S}(\mathbf{q},\omega)$, can also be accessed via quench spectroscopy, in which unequal-time correlations extracted after a sudden Hamiltonian quench yield spectral content on Fourier transformation; this approach has already been demonstrated in dipolar Rydberg XY systems~\cite{chen2025continuous}. Closely related protocols further extend the toolbox of accessible observables: Bragg spectroscopy probes density--density correlations directly in momentum and frequency space~\cite{birkl1995bragg}; minimally invasive phase-contrast (weak-measurement) imaging yields unequal-time density correlations from the cross-correlation of two time-separated weak measurements~\cite{altuntas2025weak}; partial-transfer imaging enables repeated, minimally destructive measurements of the same atomic ensemble~\cite{seroka2019partial}; and randomized-measurement (``classical-shadow'') tomography permits efficient reconstruction of a wide range of observables from a compact classical representation of the quantum state~\cite{teng2025shadow}. While some of these probes remain experimentally challenging, they are in principle achievable with current-generation hardware, and they probe excitations on top of the QSL that carry the same spectral content as the DSF reported above. Other diagnostics, such as the topological entanglement entropy or the spectra of topological excitations, also contain valuable information, but their utility across our classification is limited: the former is a universal diagnostic only at zero temperature (a condition violated by the heating during nonadiabatic state preparation~\cite{coarsening}), and the latter would require explicit access to the $e$ and $m$ sectors that lie beyond our fermionic parton framework.

Of course, strictly speaking, the mean-field wavefunction obtained from any \textit{Ansatz} should  be projected into the single-fermion-per-site subspace through numerical Gutzwiller projection methods~\cite{Wenbook} to obtain the true wavefunction in the original spin Hilbert space. Since the candidate identified here is a fully gapped $\mathbb{Z}_2$ spin liquid, its invariant gauge group is reduced from a continuous group to the discrete $\mathbb{Z}_2$, so that no gapless emergent gauge boson exists at low energies and gauge fluctuations mediate only short-ranged interactions. Together with the gapped spinon sector, this implies that the low-energy theory is not subject to singular infrared corrections: beyond-mean-field effects may renormalize nonuniversal quantities such as linewidths, overall spectral weight, or short-distance amplitudes, but they are not expected to transform the qualitative momentum-space fingerprint of one gapped $\mathbb{Z}_2$ \textit{Ansatz} into that of another. Gutzwiller projection, which enforces the local single-occupancy constraint exactly, is likewise expected to redistribute spectral weight and sharpen quantitative features rather than to modify these qualitative fingerprints. We defer a complete numerical analysis of the effects of Gutzwiller projection and gauge fluctuations to future work.

While we have focused here on a fermionic parton construction, the Rydberg $\mathbb{Z}_2$ QSL state can also be described by a bosonic theory in which fractionalized bosonic quasiparticles live on the kagome~\cite{wang2006spin} or the dice~\cite{PhysRevB.84.094419,teng2024} lattice; understanding the PSG classification of the latter is an important open question. Bringing the complementary fermionic and bosonic perspectives together---as pursued for kagome spin liquids in~\cite{Lu-Cho-Vishwanath-2017}---and systematically diagnosing the symmetry fractionalization of the $m$ particle along the lines of Ref.~\cite{Ye-Zou-2024}, which does not rely on any specific parton construction, would yield a more complete characterization of the symmetry-enriched topological order than any single-sector PSG analysis. We note, however, that in the Rydberg setting the three anyonic sectors live on three different lattices (kagome, dice, and ruby)~\cite{Samajdar-2023}, so that each such analysis would have to be carried out lattice by lattice. Finally, our work also directly paves the way to developing variational descriptions of other species of Rydberg spin liquids---for instance, with either different emergent gauge groups~\cite{giudice2022trimer,kornjavca2023trimer,bintz2024diracspinliquidquantum} or spin chirality~\cite{tarabunga2023classification}---as well as higher-spin models~\cite{homeier2024antiferromagnetic,liu2024supersoliditysimplexphasesspin1}.

\section*{Methods}

\subsection*{DMRG calculations}
The static structure factor of the Rydberg ground state shown in Fig.~\ref{fig:SF}(a) is computed using the density-matrix renormalization group (DMRG) algorithm, implemented with the ITensor library~\cite{ITensor}. We work in the regime $\Delta/\Omega = 1.7$ and $V_{ij}/\Omega = 50$ for all pairs with separation up to the third-nearest neighbor (with $V_{ij}=0$ beyond this range). The simulations are carried out on a 288-site ruby cluster placed on a finite cylindrical geometry that is periodic along the vertical direction and open along the horizontal one, and commensurate with the ruby unit cell. We retain bond dimensions up to $\chi=800$ and maintain a truncation error of less than $10^{-6}$ throughout, which is sufficient to converge the variational ground-state energy to the precision required for the observables reported in Fig.~\ref{fig:SF}(a).

\subsection*{Mean-field structure factor calculations}
For each candidate \textit{Ansatz} in Table~\ref{table:ansatze}, we diagonalize the quadratic spinon Hamiltonian in Eq.~\eqref{eq:mf_ham_singlet_triplet} and compute the static and dynamical structure factors in the fermionic spinon representation, restricting to the connected part of the spin--spin correlator. For both quantities, we focus on the longitudinal $\langle S^z_i S^z_j \rangle$ component, since measurements in the computational basis are the most convenient experimentally. The mean-field parameters are set to a uniform value on the 1NN bonds, and likewise for the 2NN bonds, with the ratio between the two chosen to be $1:r$. We then systematically tune $r$ and the strengths of the singlet and longitudinal-triplet channels for each of the 50 $\mathbb{Z}_2$ \textit{Ans\"atze}. The mean-field decoupling procedure, the diagonalization of the resulting Bogoliubov--de Gennes Hamiltonian, and the explicit expressions used to evaluate the spin structure factors are detailed in the Supplementary Information~\cite{sm}, where we also tabulate the resulting static and dynamical structure factors over a broad range of parameters for all 50 \textit{Ans\"atze}.

\section*{Data availability}
The numerical data underlying the figures of this study (DMRG ground-state structure factors and mean-field structure factors for all $\mathbb{Z}_{2}$ \textit{Ans\"atze}) have been deposited on Zenodo at [DOI to be inserted upon deposition]. Additional intermediate data, including the projective symmetry group output for each \textit{Ansatz}, are available from the corresponding author upon reasonable request.

\section*{Code availability}
The custom code used to perform the projective symmetry group classification, to construct the mean-field \textit{Ans\"atze}, and to compute the static and dynamical structure factors has been deposited on Zenodo at https://doi.org/10.5281/zenodo.20696295. The DMRG calculations were performed using the publicly available ITensor library~\cite{ITensor}.

\bibliography{references}

\section*{Acknowledgements}
We thank Andrew Daley, Subir Sachdev, Yanting Teng, and Ronny Thomale for useful discussions. This work was initiated and completed at the Aspen Center for Physics. This research was also supported in part by the Kavli Institute for Theoretical Physics (KITP) during the ``A New Spin on Quantum Magnets'' program in summer 2023. Y.~I. also acknowledges the use of the computing resources at HPCE, IIT Madras. 

\section*{Funding}
The Aspen Center for Physics is supported by National Science Foundation grant PHY-2210452 and a grant from the Simons Foundation (1161654, Troyer). The KITP is supported by NSF grant PHY-2309135. A.~M. was supported by DFG Grant No.~258499086-SFB 1170 and the W\"urzburg-Dresden Cluster of Excellence on Complexity and Topology in Quantum Matter, Grant No.~390858490-EXC 2147. Y.~I. acknowledges support from the ICTP through the Associates Programme, from the Simons Foundation through Grant No.~284558FY19, and from IIT Madras through the Institute of Eminence (IoE) program for establishing QuCenDiEM (Project No.~SP22231244CPETWOQCDHOC). R.~S. was supported by the Princeton Quantum Initiative Fellowship.

\section*{Author contributions}
R.~S. and Y.~I. designed the research. A.~M. carried out the projective symmetry group classification and the mean-field structure factor calculations, with input from Y.~I. and R.~S. R.~S. performed the DMRG simulations. All authors (A.~M., Y.~I., and R.~S.) discussed the results, interpreted the findings, and contributed to writing the manuscript.

\section*{Competing interests}
The authors declare no competing interests.

\newpage


\section*{Supplementary material}

Atanu Maity,[1, 2] Yasir Iqbal,[2] and Rhine Samajdar[3, 4]

[1]*Institut für Theoretische Physik and Würzburg-Dresden Cluster of Excellence ct.qmat,
Julius-Maximilians-Universität Würzburg, Am Hubland, Campus Süd, Würzburg 97074, Germany*
[2]*Department of Physics and Quantum Centre for Diamond and Emergent Materials (QuCenDiEM),
Indian Institute of Technology Madras, Chennai 600036, India*
[3]*Department of Physics, Princeton University, Princeton, NJ 08544, USA*
[4]*Princeton Center for Theoretical Science, Princeton, NJ 08544, USA*

In this Supplemental Material, we provide further details on the projective symmetry group (PSG) classification on the ruby lattice and the mean-field decoupling of the Rydberg Hamiltonian in the fermionic parton language. We also analytically compute and showcase the static as well as dynamical spin structure factors for all the identified *Ansätze* over a broad range of mean-field parameters.

## SI. SYMMETRIES AND PSG CLASSIFICATION

On the ruby lattice, the Rydberg Hamiltonian defined in Eq. (1) of the main text is explicitly invariant under the following symmetry transformations:

$$T_1 : (S_i^x, S_i^y, S_i^z) \rightarrow \left(S_{T_1(i)}^x, S_{T_1(i)}^y, S_{T_1(i)}^z\right), \quad (S1)$$

$$T_2 : (S_i^x, S_i^y, S_i^z) \rightarrow \left(S_{T_2(i)}^x, S_{T_2(i)}^y, S_{T_2(i)}^z\right), \quad (S2)$$

$$C_6 : (S_i^x, S_i^y, S_i^z) \rightarrow \left(S_{C_6(i)}^x, S_{C_6(i)}^y, S_{C_6(i)}^z\right), \quad (S3)$$

$$R : (S_i^x, S_i^y, S_i^z) \rightarrow \left(S_{R(i)}^x, S_{R(i)}^y, S_{R(i)}^z\right), \quad (S4)$$

$$\Theta : (S_i^x, S_i^y, S_i^z) \rightarrow (S_i^x, -S_i^y, S_i^z). \quad (S5)$$

Here, $T_1(i)$ and $T_2(i)$ denote translations of the lattice coordinate $i$ by the lattice vectors $\mathbf{T}_1$ and $\mathbf{T}_2$, respectively, $C_6(i)$ implements a sixfold rotation around the axis perpendicular to the plane, and $R(i)$ represents a reflection about the $x$-axis. Note that time reversal $\Theta$ does not act on the lattice coordinates.

In parton space, the linear actions of the space-group symmetries are straightforward since they simply modify the site indices. However, the action of $\Theta$ has to be defined more carefully. Under time reversal, we have $\Theta : \{f_{i\uparrow} \rightarrow -f_{i\downarrow}^\dagger, f_{i\downarrow} \rightarrow f_{i\uparrow}^\dagger\}$, or equivalently, on the spinor doublet, $\Theta : \psi_i \rightarrow \psi_i \, i\tau^y$. Accordingly, the spin operators transform as

$$S_i^\gamma = \frac{1}{2} \text{Tr} \left[\psi_i^\dagger \tau^\gamma \psi_i\right] \rightarrow \frac{1}{2} \text{Tr} \left[(-i\tau^y)\psi_i^\dagger (\tau^\gamma)^\star \psi_i (i\tau^y)\right]$$
$$= \frac{1}{2} \text{Tr} \left[\psi_i^\dagger (\tau^\gamma)^\star \psi_i\right] = (-1)^{\delta_{\gamma,y}} S_i^\gamma.$$

The projective extension of the symmetry group can be directly read off from Ref. [1] as:

$$W_{T_1}(x, y, s) = \eta^y \tau^0, \quad W_{T_2}(x, y, s) = \tau^0, \quad (S6)$$

$$W_{C_6}(x, y, s) = \eta^{xy + \frac{y}{2}(y+1)} \left(\eta_{C_6}\right)^{\delta_{s,6}} \tau^0, \quad (S7)$$

$$W_R(x, y, s) = \eta^{\frac{x}{2}(x+1)} \left(\eta_{C_6}\right)^{\delta_{s,5}} \left(\eta_{C_6 R}\right)^{\delta_{\text{mod}(s,2),0}} g_R, \quad (S8)$$

$$W_\Theta(x, y, s) = \eta_{\Theta C_6}^s g_\Theta, \quad g_R^2 = \eta_R \tau^0, \quad (S9)$$

where $(x, y, s)$ denotes the position of a site on sublattice $s$ in the unit cell located at $(x, y)$. In the equations above, all the parameters $\{\eta, \eta_{C_6}, \eta_{C_6 R}, \eta_{\Theta C_6}\}$ can take values $\pm 1$ and the gauge-independent choices of the matrices $g_R$ and $g_\Theta$ are listed in Table I below. In total, we obtain 64 projective symmetry extensions for a $\mathbb{Z}_2$ IGG.

| $\eta_R$ | $g_R$ | $g_\Theta$ | Set of $\eta$ parameters | # of PSGs |
|---|---|---|---|---|
| $+1$ | $\tau^0$ | $i\tau^y$ | $\{\eta_{\Theta C_6}, \eta, \eta_{C_6}, \eta_{C_6 R}\}$ | $2^4$ |
| $+1$ | $\tau^0$ | $\tau^0$ | $\{\eta_{\Theta C_6} = -1, \eta, \eta_{C_6}, \eta_{C_6 R}\}$ | $2^3$ |
| $-1$ | $i\tau^z$ | $i\tau^y$ | $\{\eta_{\Theta C_6}, \eta, \eta_{C_6}, \eta_{C_6 R}\}$ | $2^4$ |
| $-1$ | $i\tau^z$ | $i\tau^z$ | $\{\eta_{\Theta C_6}, \eta, \eta_{C_6}, \eta_{C_6 R}\}$ | $2^4$ |
| $-1$ | $i\tau^z$ | $\tau^0$ | $\{\eta_{\Theta C_6} = -1, \eta, \eta_{C_6}, \eta_{C_6 R}\}$ | $2^3$ |

TABLE I. All possible gauge-inequivalent choices of the binary parameter $\eta_R$, and the two matrices $g_R, g_\Theta$, which together define a total of 64 $\mathbb{Z}_2$ PSG solutions.

Once the PSGs have been specified, we can proceed to construct the mean-field *Ansätze*. In our projective construction, the symmetries of the mean fields are preserved as

$$u_{ij}^0 = (-1)^{\delta_{\mathcal{O},\Theta}} W_{\mathcal{O}}^\dagger(\mathcal{O}(i)) u_{\mathcal{O}(i)\mathcal{O}(j)}^0 W_{\mathcal{O}}(\mathcal{O}(j)),$$

$$u_{ij}^{\gamma'} = W_{\mathcal{O}}^\dagger(\mathcal{O}(i)) u_{\mathcal{O}(i)\mathcal{O}(j)}^\gamma \mathcal{R}_{\mathcal{O}}^{\gamma\gamma'} W_{\mathcal{O}}(\mathcal{O}(j)), \quad (S10)$$

where $\gamma, \gamma' = \{x, y, z\}$ and $\mathcal{R}_{\mathcal{O}}$ are $3 \times 3$ matrices. These matrices are, in general, nontrivial for real spin operators which transform as pseudovectors under symmetry operations. In our case, however, for the lattice space-group symmetry elements, $\mathcal{R}_{\mathcal{O}} = \mathbb{1}_{3\times 3}$.

Now, let us consider the effect of time reversal on the *Ansatz*. For singlet fields, i.e., $u_{ij}^0 = is_{ij}^0\tau^0 + s_{ij}^x\tau^x + $

$s_{ij}^2 \tau^y + s_{ij}^3 \tau^z$, we have

$$\Theta : \mathrm{Tr}\left[\psi_i u_{ij}^0 \psi_j^\dagger\right] \to \mathrm{Tr}\left[\psi_i (i\tau^y)(u_{ij}^0)^\star(-i\tau^y)\psi_j^\dagger\right]$$
$$= -\mathrm{Tr}\left[\psi_i u_{ij}^0 \psi_j^\dagger\right], \tag{S11}$$

whereas for the triplet components, i.e., $u_{ij}^\gamma = t_{ij}^{\gamma,0}\tau^0 + i(t_{ij}^{\gamma,1}\tau^x + t_{ij}^{\gamma,2}\tau^y + t_{ij}^{\gamma,3}\tau^z)$,

$$\Theta : \mathrm{Tr}\left[\tau^\gamma \psi_i u_{ij}^\gamma \psi_j^\dagger\right] \to \mathrm{Tr}\left[(\tau^\gamma)^\star \psi_i (i\tau^y)(u_{ij}^\gamma)^\star(-i\tau^y)\psi_j^\dagger\right]$$
$$= (-)^{\delta_{\gamma,y}} \mathrm{Tr}\left[\tau^\gamma \psi_i u_{ij}^\gamma \psi_j^\dagger\right]. \tag{S12}$$

Therefore, we obtain

$$\Theta : u_{ij}^0 \to -u_{ij}^0, \tag{S13}$$
$$\Theta : \{u_{ij}^x, u_{ij}^y, u_{ij}^z\} \to \{u_{ij}^x, -u_{ij}^y, u_{ij}^z\}, \tag{S14}$$

so $\mathcal{R}_\Theta = \mathrm{Diag}[1,-1,1]$ with an additional negative sign for the singlet fields.

As mentioned in the main text, the full spatial structure of the *Ansatz* can be defined by specifying a few reference mean fields. Temporarily using the notation $u_{s_1,s_2}^\alpha$ ($s_1, s_2 = 1, \ldots, 6$) to denote the directed link *from* a site on sublattice $s_1$ *to* a site on sublattice $s_2$, the sign structure of the various terms is given by

$$u_{1,3}^\alpha = u_{2,4}^\alpha = \eta u_{3,5}^\alpha = \eta\eta_{C_6} u_{4,6}^\alpha = \eta\eta_{C_6} u_{5,1}^\alpha = \eta u_{6,2}^\alpha \equiv \mathcal{U}_1^\alpha, \tag{S15}$$

$$u_{1,2}^\alpha = u_{2,3}^\alpha = u_{3,4}^\alpha = u_{4,5}^\alpha = \eta_{C_6} u_{5,6}^\alpha = u_{6,1}^\alpha \equiv \mathcal{U}_2^\alpha, \tag{S16}$$

$$u_{1,4}^\alpha = \eta\eta_{C_6} u_{3,6}^\alpha = \eta\eta_{C_6} u_{5,2}^\alpha \equiv \mathcal{U}_3^\alpha, \tag{S17}$$

$$u_{2,5}^\alpha = \eta\eta_{C_6} u_{4,1}^\alpha = \eta u_{6,3}^\alpha = \eta\eta_{C_6}\eta_{C_6R} g_R^\dagger \mathcal{U}_3^\alpha g_R, \tag{S18}$$

where the subscript 1, 2, or 3 on $\mathcal{U}^\alpha$ indicates the first-, second-, or third-nearest-neighbor nature of the bond. Furthermore, the bonds connecting sites with different $y$-coordinates also change sign under translations by $\mathbf{T}_1$.

## SII. MEAN-FIELD DECOUPLING

In the Abrikosov fermion representation $S_i^\gamma = \frac{1}{2} f_{i\sigma}^\dagger \tau_{\sigma\sigma'}^\gamma f_{i\sigma'}$, the Rydberg Hamiltonian can be recast as $H = H_0 + H_2 + H_4$, where

$$H_0 = -\frac{\Delta N}{2} + \sum_{\langle i,j \rangle} \frac{V_{ij}}{4}, \tag{S19}$$

$$H_2 = \frac{1}{2}\sum_i \Omega f_{i,\sigma}^\dagger \tau_{\sigma\sigma'}^x f_{i,\sigma'} - \frac{1}{2}\sum_i \Delta f_{i,\sigma}^\dagger \tau_{\sigma\sigma'}^z f_{i,\sigma'} + \sum_{\langle i,j \rangle} \frac{V_{ij}}{4}\left[f_{i,\sigma}^\dagger \tau_{\sigma\sigma'}^z f_{i,\sigma'} + f_{j,\sigma}^\dagger \tau_{\sigma\sigma'}^z f_{j,\sigma'}\right], \tag{S20}$$

$$H_4 = \frac{1}{4}\sum_{\langle i,j \rangle} V_{ij}\, \sigma\sigma' f_{j\sigma'}^\dagger f_{i\sigma}^\dagger f_{i\sigma} f_{j\sigma'}. \tag{S21}$$

Now, we perform a quadratic decoupling of the quartic term to obtain a mean-field theory as

$$\sigma\sigma' f_{j\sigma'}^\dagger f_{i\sigma}^\dagger f_{i\sigma} f_{j\sigma'} = \left\langle f_{i\sigma}^\dagger f_{j\sigma}^\dagger \right\rangle f_{i\sigma} f_{j\sigma} + \left\langle f_{i\sigma} f_{j\sigma} \right\rangle f_{i\sigma}^\dagger f_{j\sigma}^\dagger - \left\langle f_{i\sigma}^\dagger f_{j\sigma}^\dagger \right\rangle \left\langle f_{i\sigma} f_{j\sigma} \right\rangle$$
$$- \left\langle f_{i\sigma}^\dagger f_{j\bar\sigma}^\dagger \right\rangle f_{i\sigma} f_{j\bar\sigma} - \left\langle f_{i\sigma} f_{j\bar\sigma} \right\rangle f_{i\sigma}^\dagger f_{j\bar\sigma}^\dagger + \left\langle f_{i\sigma}^\dagger f_{j\bar\sigma}^\dagger \right\rangle \left\langle f_{i\sigma} f_{j\bar\sigma} \right\rangle$$
$$- \left\langle f_{i\sigma}^\dagger f_{j\sigma} \right\rangle f_{j\sigma'}^\dagger f_{i\sigma} - \left\langle f_{j\sigma'}^\dagger f_{i\sigma} \right\rangle f_{i\sigma}^\dagger f_{j\sigma} + \left\langle f_{i\sigma}^\dagger f_{j\sigma} \right\rangle \left\langle f_{j\sigma'}^\dagger f_{i\sigma} \right\rangle$$
$$+ \left\langle f_{i\sigma}^\dagger f_{j\bar\sigma} \right\rangle f_{j\bar\sigma}^\dagger f_{i\sigma} + \left\langle f_{j\bar\sigma}^\dagger f_{i\sigma} \right\rangle f_{i\sigma}^\dagger f_{j\bar\sigma} - \left\langle f_{i\sigma}^\dagger f_{j\bar\sigma} \right\rangle \left\langle f_{j\bar\sigma}^\dagger f_{i\sigma} \right\rangle$$
$$+ \left\langle \sigma f_{i\sigma}^\dagger f_{i\sigma} \right\rangle \sigma' f_{j\sigma'}^\dagger f_{j\sigma'} + \left\langle \sigma' f_{j\sigma'}^\dagger f_{j\sigma'} \right\rangle \sigma f_{i\sigma}^\dagger f_{i\sigma} - \left\langle \sigma f_{i\sigma}^\dagger f_{i\sigma} \right\rangle \left\langle \sigma' f_{j\sigma'}^\dagger f_{j\sigma'} \right\rangle. \tag{S22}$$

Next, we define the following singlet and triplet hopping and pairing fields:

$$\chi_{ij} = f_{i\uparrow}^\dagger f_{j\uparrow} + f_{i\downarrow}^\dagger f_{j\downarrow}, \qquad \xi_{ij} = f_{i\uparrow}^\dagger f_{j\downarrow}^\dagger - f_{i\downarrow}^\dagger f_{j\uparrow}^\dagger, \tag{S23}$$

$$T_{ij}^{x,h} = f_{i\uparrow}^\dagger f_{j\downarrow} + f_{i\downarrow}^\dagger f_{j\uparrow}, \quad T_{ij}^{y,h} = -i(f_{i\uparrow}^\dagger f_{j\downarrow} - f_{i\downarrow}^\dagger f_{j\uparrow}), \quad T_{ij}^{z,h} = f_{i\uparrow}^\dagger f_{j\uparrow} - f_{i\downarrow}^\dagger f_{j\downarrow}, \tag{S24}$$

$$T_{ij}^{x,p} = f_{i\downarrow}^\dagger f_{j\downarrow}^\dagger - f_{i\uparrow}^\dagger f_{j\uparrow}^\dagger, \quad T_{ij}^{y,p} = i(f_{i\downarrow}^\dagger f_{j\downarrow}^\dagger + f_{i\uparrow}^\dagger f_{j\uparrow}^\dagger), \qquad T_{ij}^{z,p} = f_{i\uparrow}^\dagger f_{j\downarrow}^\dagger + f_{i\downarrow}^\dagger f_{j\uparrow}^\dagger. \tag{S25}$$

In the mean-field approximation, i.e., $\langle \chi_{ij} \rangle = \chi_{ij}$, $\langle \xi_{ij} \rangle = \xi_{ij}$, and $\langle T_{ij}^{\gamma,r} \rangle = T_{ij}^{\gamma,r}$ (with $r = h, p$), the Hamiltonian reads

$$H = H_0 + H_2 + H_{\tilde{2}}, \tag{S26}$$

where

$$H_{\tilde{0}} = H_0 + \frac{1}{8} \sum_{\langle i,j \rangle} V_{ij} \left[ |\chi_{ij}|^2 + |\xi_{ij}|^2 + \frac{1}{8} \sum_{\gamma,r} \zeta_\gamma |T_{ij}^{\gamma,r}|^2 - 2T_{ii}^{z,h} T_{jj}^{z,h} \right], \tag{S27}$$

$$H_2 = \frac{1}{2} \sum_i \Omega\, T_{ii}^{x,h} - \frac{1}{2} \sum_i \Delta T_{ii}^{z,h} + \sum_{\langle i,j \rangle} \frac{V_{ij}}{4} \left[ T_{ii}^{z,h} + T_{jj}^{z,h} \right], \tag{S28}$$

$$H_{\tilde{2}} = -\frac{1}{8} \sum_{\langle i,j \rangle} V_{ij} \left[ \left( \chi_{ij}^\star \chi_{ij} + \xi_{ij}^\star \xi_{ij} + \sum_{\gamma,r} \zeta_\gamma (T_{ij}^{\gamma,r})^\star T_{ij}^{\gamma,r} + \text{h.c.} \right) - 2T_{ii}^{z,h} T_{jj}^{z,h} - 2T_{jj}^{z,h} T_{ii}^{z,h} \right], \tag{S29}$$

with $\zeta_z = -\zeta_x = -\zeta_y = +1$. Now, using the doublet $\psi_i$ and the matrices

$$u_{ij}^0 = \begin{bmatrix} \chi_{ij} & \xi_{ij} \\ \xi_{ij}^\star & -\chi_{ij}^\star \end{bmatrix}, \quad u_{ij}^\gamma = \begin{bmatrix} T_{ij}^{\gamma,h} & T_{ij}^{\gamma,p} \\ -(T_{ij}^{\gamma,p})^\star & (T_{ij}^{\gamma,h})^\star \end{bmatrix}, \tag{S30}$$

one can rewrite the quadratic part, i.e., $H_q = H_2 + H_{\tilde{2}}$, in the form

$$H_q = \frac{1}{16} \sum_{\langle i,j \rangle} V_{ij} \left( \text{Tr}[\tau^\alpha \psi_i\, u_{ij}^\alpha \psi_j^\dagger] + \text{h.c.}] + \text{Tr}[\tau^z \psi_i\, u_{ii}^z \psi_i^\dagger] + \text{Tr}[\tau^z \psi_j\, u_{jj}^z \psi_j^\dagger] \right) + \frac{1}{4} \sum_i \left( \Omega\, \text{Tr}[\tau^x \psi_i\, \psi_i^\dagger] - \Delta\, \text{Tr}[\tau^z \psi_i\, \psi_i^\dagger] \right), \tag{S31}$$

where $u_{ii}^z$ is a $2 \times 2$ diagonal matrix with elements $\tilde{T}_{ii}^{z,h}$ and $(\tilde{T}_{ii}^{z,h})^\dagger$; $\tilde{T}_{ii}^{z,h} \equiv 2T_{ii}^{z,h} + 1$. If we also restrict the Hamiltonian to $m^{th}$-nearest-neighbor ($m$NN) terms and there are, say, $N_p$ $p$NN sites per unit cell, then we have

$$\begin{aligned} H_q &= \frac{1}{16} \sum_{p=1}^m \sum_{\langle i,j \rangle_p} V_p \left( \text{Tr}[\tau^\alpha \psi_i\, u_{ij}^\alpha \psi_j^\dagger] + \text{h.c.}] + N_p \text{Tr}[\tau^z \psi_i\, u_{ii}^z \psi_i^\dagger] \right) + \frac{1}{4} \sum_i \left( \Omega\, \text{Tr}[\tau^x \psi_i\, \psi_i^\dagger] - \Delta\, \text{Tr}[\tau^z \psi_i\, \psi_i^\dagger] \right) \\ &= \frac{1}{16} \sum_{p=1}^m \sum_{\langle i,j \rangle_p} V_p \left( \text{Tr}[\tau^\alpha \psi_i\, u_{ij}^\alpha \psi_j^\dagger] + \text{h.c.}] \right) + \frac{1}{4} \sum_i \left( \Omega\, \text{Tr}[\tau^x \psi_i\, \psi_i^\dagger] + \text{Tr}[\tau^z \psi_i\, \tilde{u}_{ii}^z \psi_i^\dagger] \right), \end{aligned} \tag{S32}$$

where we use the shorthand $\tilde{u}_{ii}^z \equiv -\Delta \tau^0 + u_{ii}^z \sum_{p=1}^m N_p V_p/8$, and $V_p$ is the van der Waals interaction between $p^{th}$-neighboring sites. Note that each $u_{ij}^\alpha$ consists of a hopping and a pairing field which are, in general, complex. Additionally, we denote $u_{ii}^x = \Omega \tau^0$, rename $\tilde{u}_{ij}^z$ as $u_{ii}^z$, and absorb the constants by rescaling the $u$ fields. This sequence of steps then finally brings us to expression for the quadratic Hamiltonian in Eq. (2) of the main text:

$$H_q = \sum_{p=1}^m \sum_{\langle i,j \rangle_p} V_p\, \text{Tr} \left[ \tau^\alpha \psi_i\, u_{ij}^\alpha \psi_j^\dagger + \text{h.c.} \right] + \sum_{i,\alpha=\{x,y,z\}} \text{Tr} \left[ \tau^\alpha \psi_i\, u_{ii}^\alpha \psi_i^\dagger \right]. \tag{S33}$$

## SIII. SPIN STRUCTURE FACTORS

In this section, we first outline the procedure to calculate the spin structure factors in the fermionic spinon representation. Considering only the longitudinal component, the dynamical structure factor is defined as

$$\mathcal{S}(\mathbf{q}, \omega) = \int_{-\infty}^{+\infty} \frac{dt}{2\pi N^2} e^{i\omega t} \sum_{i,j} e^{i\mathbf{q}\cdot\mathbf{r}_{ij}} \left( \langle S_i^z(t) S_j^z(0) \rangle - \langle S_i^z(t) \rangle \langle S_j^z(0) \rangle \right) \equiv \mathcal{S}_{\text{tot}}(\mathbf{q}, \omega) - \mathcal{S}_{\text{dis}}(\mathbf{q}, \omega), \tag{S34}$$

where $\mathbf{r}_{ij} \equiv \mathbf{r}_i - \mathbf{r}_j$ and $N$ is the total number of spins. In this definition, the connected correlator $\mathcal{S}(\mathbf{q}, \omega)$ is interpreted as the sum of two pieces: $\mathcal{S}_{\text{tot}}(\mathbf{q}, \omega)$, which is the "total" correlation function $\langle S_i^z(t) S_j^z(0) \rangle$, and a disconnected part, $\mathcal{S}_{\text{dis}}(\mathbf{q}, \omega)$. Later, we will show how the disconnected contribution can be exactly canceled out by considering only the connected scattering mechanism.

To begin, substituting $S_i^z(t) = e^{iHt} S_i^z e^{-iHt}$ in terms of the fermion operators, $\mathcal{S}_{\text{tot}}(\mathbf{q}, \omega)$ is easily reformulated to

$$\mathcal{S}_{\text{tot}}(\mathbf{q}, \omega) = \int_{-\infty}^{+\infty} \frac{dt}{8\pi N^2} e^{i\omega t} \sum_{i,j} e^{i\mathbf{q}\cdot\mathbf{r}_{ij}} \tau_{\sigma\sigma}^z \tau_{\sigma'\sigma'}^z \sum_{\sigma,\sigma'} \left\langle e^{iHt} f_{i,\sigma}^\dagger f_{i,\sigma} e^{-iHt} f_{j,\sigma'}^\dagger f_{j,\sigma'} \right\rangle. \tag{S35}$$

Before proceeding further, however, let us examine the generic structure of the quadratic spinon Hamiltonian. To do so, we consider the Bogoliubov–de Gennes(BdG) basis:

$$\psi = \left( \psi_\uparrow, \psi_\downarrow, (\psi_\uparrow^\dagger)^{\text{T}}, (\psi_\downarrow^\dagger)^{\text{T}} \right)^{\text{T}} ; \ \psi_\sigma = \left( f_{1,\sigma}, f_{2,\sigma}, \cdots, f_{N,\sigma} \right). \tag{S36}$$

Any general mean-field Hamiltonian can always be expressed in the form $H = \psi^\dagger H_{\text{BdG}} \psi$. The eigenvalues come in positive and negative pairs ($\pm\epsilon_{\mu,\sigma}$) and, in the diagonal basis (obtained via a $4N \times 4N$ unitary transformation $U$),

$$\phi = \left( \phi_\uparrow, \phi_\downarrow, (\phi_\uparrow^\dagger)^{\text{T}}, (\phi_\downarrow^\dagger)^{\text{T}} \right)^{\text{T}}, \ \psi_\sigma = \left( c_{1,\sigma}, c_{2,\sigma}, \cdots, c_{N,\sigma} \right), \tag{S37}$$

the BdG Hamiltonian takes the form

$$H_{\text{BdG}}^d = U^\dagger H_{\text{BdG}} U = \sum_{\mu=1}^N \sum_{\sigma=\uparrow,\downarrow} \epsilon_{\mu,\sigma} \left( c_{\mu,\sigma}^\dagger c_{\mu,\sigma} - c_{\mu,\sigma} c_{\mu,\sigma}^\dagger \right) = \sum_{\mu=1}^N \sum_{\sigma=\uparrow,\downarrow} \epsilon_{\mu,\sigma} \left( 2c_{\mu,\sigma}^\dagger c_{\mu,\sigma} - 1 \right). \tag{S38}$$

This construction permits us to straightforwardly interpret the Bogoliubov quasiparticles as excitations with positive energy. Thus, the Fermi level lies at zero energy, and the ground state has no Bogoliubov excitations, by definition. The associated energy eigenvalues can also be listed as

$$\left( \cdots, \epsilon_{m,\uparrow}, \cdots, \epsilon_{m,\downarrow}, \cdots, \epsilon_{m,\uparrow}, \cdots, \epsilon_{m,\downarrow}, \cdots \right) \equiv \left( \cdots, \epsilon_m, \cdots, \epsilon_{m+N}, \cdots, \epsilon_{m+2N}, \cdots, \epsilon_{m+3N}, \cdots \right), \tag{S39}$$

and this is the notation we use hereafter. In terms of the new $c$-fermions, the original $f$-fermion operators are given by

$$\begin{aligned}
f_{i,\uparrow} &= \sum_{\mu=1}^N \left( U_{i,\mu} c_{\mu,\uparrow} + U_{i,\mu+N} c_{\mu,\downarrow} + U_{i,\mu+2N} c_{\mu,\uparrow}^\dagger + U_{i,\mu+3N} c_{\mu,\downarrow}^\dagger \right), \\
f_{i,\downarrow} &= \sum_{\mu=1}^N \left( U_{i+N,\mu} c_{\mu,\uparrow} + U_{i+N,\mu+N} c_{\mu,\downarrow} + U_{i+N,\mu+2N} c_{\mu,\uparrow}^\dagger + U_{i+N,\mu+3N} c_{\mu,\downarrow}^\dagger \right), \\
f_{i,\uparrow}^\dagger &= \sum_{\mu=1}^N \left( U_{i+2N,\mu} c_{\mu,\uparrow} + U_{i+2N,\mu+N} c_{\mu,\downarrow} + U_{i+2N,\mu+2N} c_{\mu,\uparrow}^\dagger + U_{i+2N,\mu+3N} c_{\mu,\downarrow}^\dagger \right), \\
f_{i,\downarrow}^\dagger &= \sum_{\mu=1}^N \left( U_{i+3N,\mu} c_{\mu,\uparrow} + U_{i+3N,\mu+N} c_{\mu,\downarrow} + U_{i+3N,\mu+2N} c_{\mu,\uparrow}^\dagger + U_{i+3N,\mu+3N} c_{\mu,\downarrow}^\dagger \right).
\end{aligned} \tag{S40}$$

This correspondence, as we now demonstrate, enables us to build up the spin structure factors piece-by-piece.

Consider first the term $\mathcal{T}_{\sigma\sigma'} \equiv \left\langle e^{iHt} f_{i,\sigma}^\dagger f_{i,\sigma} e^{-iHt} f_{j,\sigma'}^\dagger f_{j,\sigma'} \right\rangle$ with $\sigma = \sigma' = \uparrow$. Using Eq. (S40), one can write

$$\begin{aligned}
\mathcal{T}_{\uparrow\uparrow} = \sum_{\mu,\nu,\rho,\lambda=1}^N \Big\langle 0 \Big| & e^{iHt} \left( U_{i,\rho}^\star c_{\rho,\uparrow}^\dagger + U_{i,\rho+N}^\star c_{\rho,\downarrow}^\dagger + U_{i,\rho+2N}^\star c_{\rho,\uparrow} + U_{i,\rho+3N}^\star c_{\rho,\downarrow} \right) \\
& \times \left( U_{i,\lambda} c_{\lambda,\uparrow} + U_{i,\lambda+N} c_{\lambda,\downarrow} + U_{i,\lambda+2N} c_{\lambda,\uparrow}^\dagger + U_{i,\lambda+3N} c_{\lambda,\downarrow}^\dagger \right) e^{-iHt} \\
& \times \left( U_{j,\mu}^\star c_{\mu,\uparrow}^\dagger + U_{j,\mu+N}^\star c_{\mu,\downarrow}^\dagger + U_{j,\mu+2N}^\star c_{\mu,\uparrow} + U_{j,\mu+3N}^\star c_{\mu,\downarrow} \right) \\
& \times \left( U_{j,\nu} c_{\nu,\uparrow} + U_{j,\nu+N} c_{\nu,\downarrow} + U_{j,\nu+2N} c_{\nu,\uparrow}^\dagger + U_{j,\nu+3N} c_{\nu,\downarrow}^\dagger \right) \Big| 0 \Big\rangle.
\end{aligned} \tag{S41}$$

Recalling that the ground state is the Bogoliubov vacuum $|0\rangle$, a scattering process follows the creation of a pair of Bogoliubov quasiparticles at time $0$ and their annihilation at a later time $t$. Therefore, the terms of the form

"$c_{\rho,\sigma_1}(t)c_{\lambda,\sigma_2}(t)c_{\mu,\sigma_1'}^\dagger(0)c_{\nu,\sigma_2'}^\dagger(0)$" *will* contribute to the structure factors. On the other hand, this ensures the exclusion of the disconnected contributions, i.e., $\mathcal{S}_{\mathrm{dis}}(\mathbf{q},\omega)$ in Eq. (S34), arising from terms like "$c_{\rho,\sigma_1}(t)c_{\rho,\sigma_1}^\dagger(t)c_{\mu,\sigma_1'}c_{\mu,\sigma_1'}^\dagger(0)$" where the process follows the creation of quasiparticles $(\mu,\sigma_1')$ and $(\rho,\sigma_1)$ at times 0 and $t$, respectively, followed by their annihilation at the same time.

The connected pieces, which we concentrate on, can be further simplified. To illustrate this, let us take, for example, the term $\langle 0|e^{iHt}c_{\rho,\uparrow}c_{\lambda,\uparrow}e^{-iHt}c_{\mu,\uparrow}^\dagger c_{\nu,\uparrow}^\dagger|0\rangle$, which contributes to the DSF as

$$
\begin{aligned}
&\int_{-\infty}^{+\infty}\frac{dt}{8\pi N^2}e^{i\omega t}\sum_{i,j}e^{i\mathbf{q}\cdot\mathbf{r}_{ij}}\sum_{\mu,\nu,\rho,\lambda}U_{i,\rho+2N}^\star U_{i,\lambda}U_{j,\mu}^\star U_{j,\nu+2N}\langle 0|e^{iHt}c_{\rho,\uparrow}c_{\lambda,\uparrow}e^{-iHt}c_{\mu,\uparrow}^\dagger c_{\nu,\uparrow}^\dagger|0\rangle \\
&=\int_{-\infty}^{+\infty}\frac{dt}{8\pi N^2}e^{i\omega t}\sum_{i,j}e^{i\mathbf{q}\cdot\mathbf{r}_{ij}}\sum_{\mu,\nu,\rho,\lambda}U_{i,\rho+2N}^\star U_{i,\lambda}U_{j,\mu}^\star U_{j,\nu+2N}\langle 0|c_{\rho,\uparrow}c_{\lambda,\uparrow}c_{\mu,\uparrow}^\dagger c_{\nu,\uparrow}^\dagger|0\rangle e^{-i(\epsilon_\mu+\epsilon_{\nu+2N})t} \\
&=\frac{1}{4N^2}\sum_{i,j}e^{i\mathbf{q}\cdot\mathbf{r}_{ij}}\sum_{\mu,\nu,\rho,\lambda}U_{i,\rho+2N}^\star U_{i,\lambda}U_{j,\mu}^\star U_{j,\nu+2N}\langle 0|c_{\rho,\uparrow}c_{\lambda,\uparrow}c_{\mu,\uparrow}^\dagger c_{\nu,\uparrow}^\dagger|0\rangle\delta(\epsilon_\mu+\epsilon_{\nu+2N}-\omega) \\
&=\frac{1}{4N^2}\sum_{i,j}e^{i\mathbf{q}\cdot\mathbf{r}_{ij}}\sum_{\mu,\nu,\rho,\lambda}U_{i,\rho+2N}^\star U_{i,\lambda}U_{j,\mu}^\star U_{j,\nu+2N}(\delta_{\rho,\nu}\delta_{\lambda,\mu}-\delta_{\rho,\mu}\delta_{\lambda,\nu})\delta(\epsilon_\mu+\epsilon_{\nu+2N}-\omega) \\
&=\frac{1}{4N^2}\sum_{i,j,\mu,\nu}e^{i\mathbf{q}\cdot\mathbf{r}_{ij}}(U_{i,\nu+2N}^\star U_{i,\mu}-U_{i,\mu+2N}^\star U_{i,\nu})U_{j,\mu}^\star U_{j,\nu+2N}\delta(\epsilon_\mu+\epsilon_{\nu+2N}-\omega).
\end{aligned}
\tag{S42}
$$

One can analogously derive all the other terms constituting $\mathcal{T}_{\uparrow\uparrow}$. Combining these, the net contribution of $\mathcal{T}_{\uparrow\uparrow}$ to $\mathcal{S}(\mathbf{q},\omega)$ takes the following form:

$$
\begin{aligned}
(\mathcal{S}(\mathbf{q},\omega))_{\uparrow\uparrow}=\frac{1}{4N^2}\sum_{i,j}e^{i\mathbf{q}\cdot\mathbf{r}_{ij}}\sum_{\mu,\nu}\Big[&(U_{i,\nu+2N}^\star U_{i,\mu}-U_{i,\mu+2N}^\star U_{i,\nu})U_{j,\mu}^\star U_{j,\nu+2N}\delta(\epsilon_\mu+\epsilon_{\nu+2N}-\omega) \\
&+(U_{i,\nu+3N}^\star U_{i,\mu+N}-U_{i,\mu+3N}^\star U_{i,\nu+N})U_{j,\mu+N}^\star U_{j,\nu+3N}\delta(\epsilon_{\mu+N}+\epsilon_{\nu+3N}-\omega) \\
&+(U_{i,\nu+2N}^\star U_{i,\mu+N}-U_{i,\mu+3N}^\star U_{i,\nu})U_{j,\mu+N}^\star U_{j,\nu+2N}\delta(\epsilon_{\mu+N}+\epsilon_{\nu+2N}-\omega) \\
&+(U_{i,\nu+3N}^\star U_{i,\mu}-U_{i,\mu+2N}^\star U_{i,\nu+N})U_{j,\mu}^\star U_{j,\nu+3N}\delta(\epsilon_\mu+\epsilon_{\nu+3N}-\omega)\Big] \\
\equiv\frac{1}{4N^2}\sum_{i,j}e^{i\mathbf{q}\cdot\mathbf{r}_{ij}}g(i,j).&
\end{aligned}
\tag{S43}
$$

Carrying out a similar analysis for the other components, i.e., $(\mathcal{S}(\mathbf{q},\omega))_{\downarrow\downarrow}$, $(\mathcal{S}(\mathbf{q},\omega))_{\uparrow\downarrow}$, and $(\mathcal{S}(\mathbf{q},\omega))_{\downarrow\uparrow}$, one finds the full expression for the DSF to be

$$
\mathcal{S}(\mathbf{q},\omega)=\frac{1}{4N^2}\sum_{i,j}e^{i\mathbf{q}\cdot\mathbf{r}_{ij}}[g(i,j)+g(i+N,j+N)-g(i,j+N)-g(i+N,j)].
\tag{S44}
$$

The static structure factor, which is the equal-time momentum-resolved spin-spin correlation function, can then directly be calculated from this equation as $S(\mathbf{q})=\sum_\omega\mathcal{S}(\mathbf{q},\omega)$.

Following the procedure outlined above, we systematically compute the static and dynamical structure factors for all our $\mathbb{Z}_2$ QSL *Ansätze*, over a broad range of parameters, and these are presented in Figs. S1 through S6, and Figs. S7 to S12, respectively.

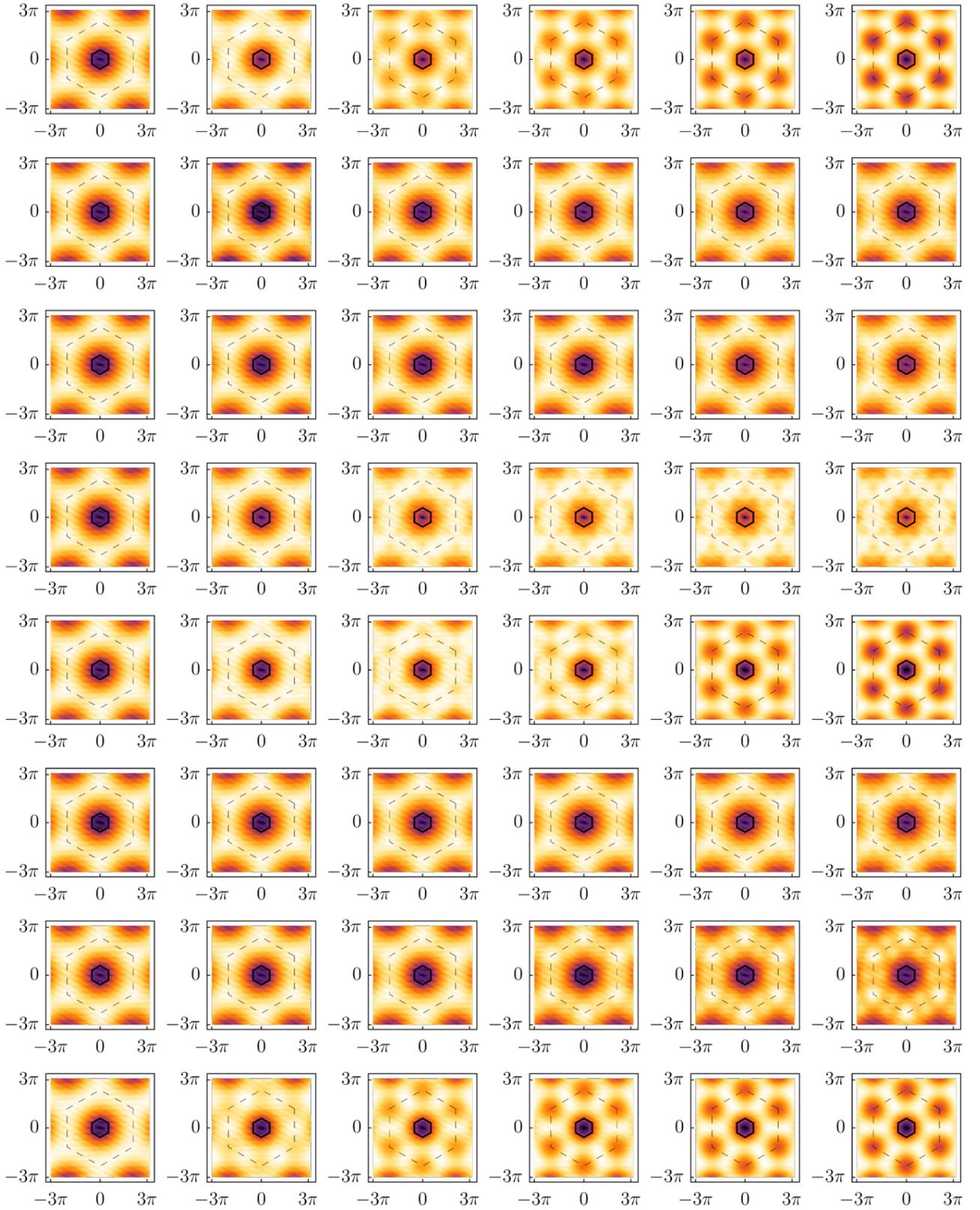

FIG. S1. Static structure factors for the $\mathbb{Z}_2$ *Ansätze* listed in the first eight rows of Table I, with $\eta_{C_6} = \eta = +1$, for a system size of $10 \times 10 \times 6$ sites. The mean-field parameters for the first-nearest-neighbor (1NN) and second-nearest-neighbor (2NN) bonds are chosen in the ratio $1 : r$, while those on the third-nearest-neighbor (3NN) bonds are set to zero. The six columns, from left to right, correspond to $r = 0.25$, $0.50$, $0.75$, $1.00$, $1.25$, and $1.50$, respectively. Without loss of generality, we choose an onsite triplet term of magnitude $0.2$. The solid (dashed) hexagon marks the first (extended) Brillouin zone.

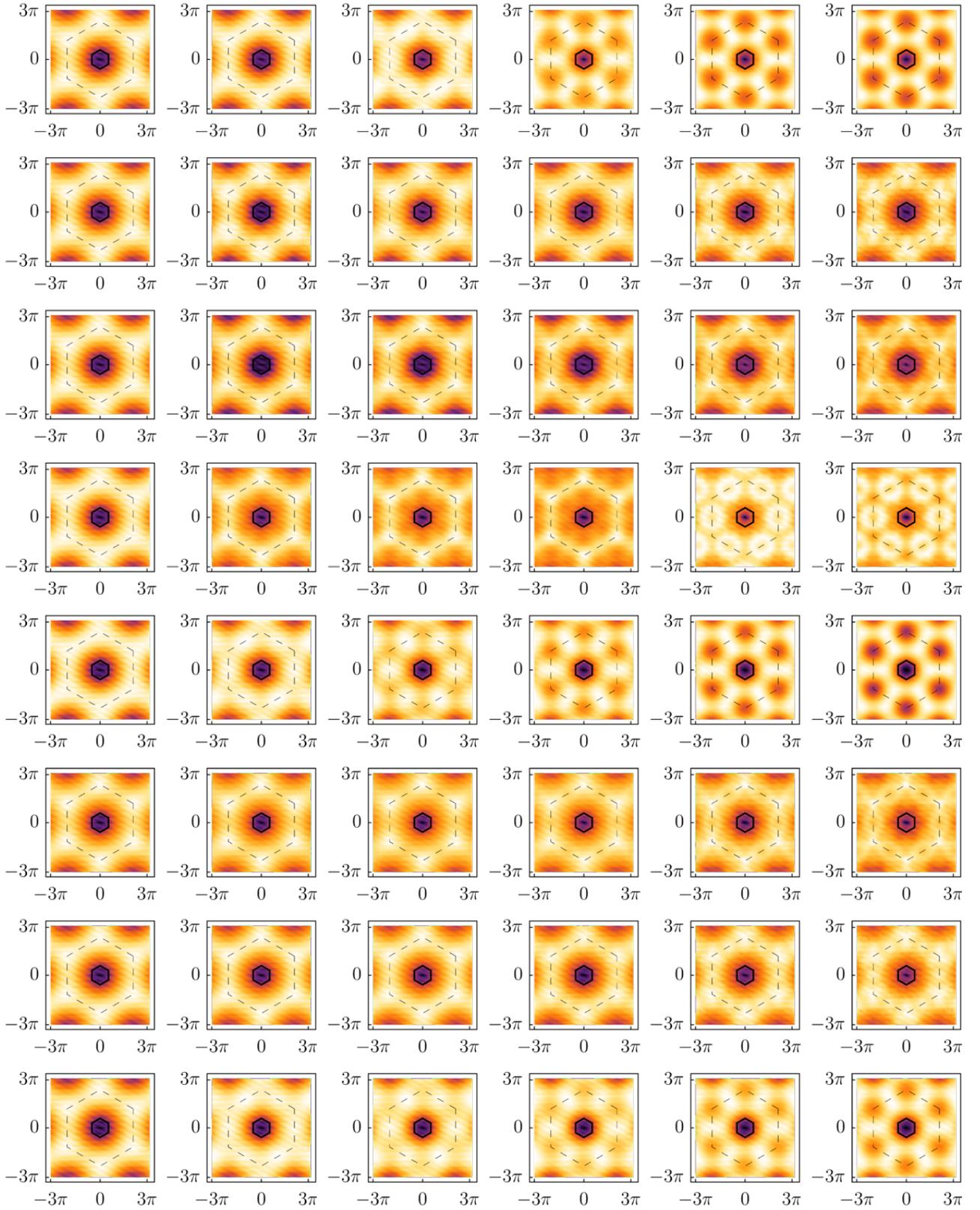

FIG. S2. Same as in Fig. S1 but for the $\mathbb{Z}_2$ *Ansätze* in the $\eta_{C_6} = \eta = -1$ class.

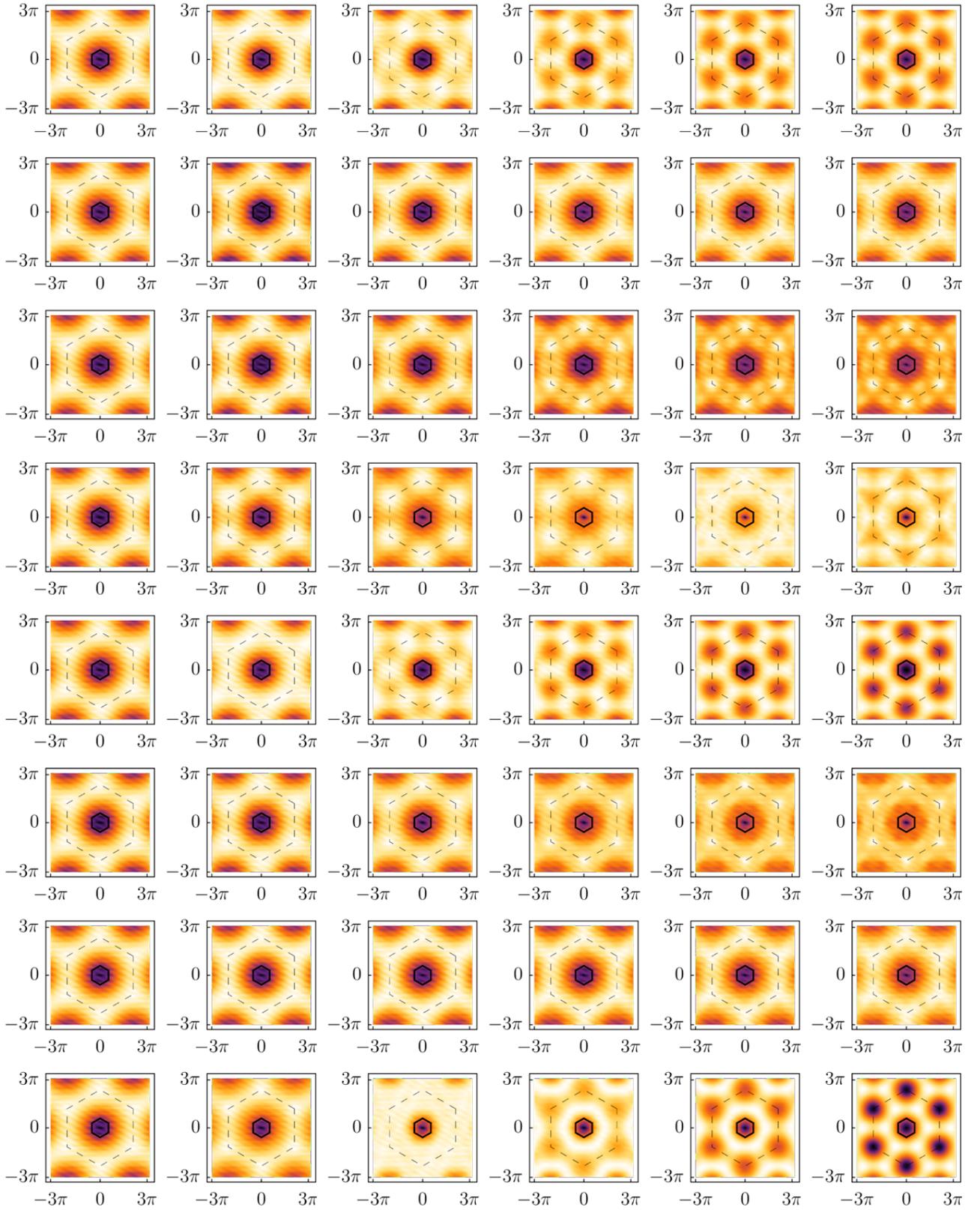

FIG. S3. Same as in Fig. S1 but for the $\mathbb{Z}_2$ *Ansätze* in the $\eta_{C_6} = -\eta = -1$ class.

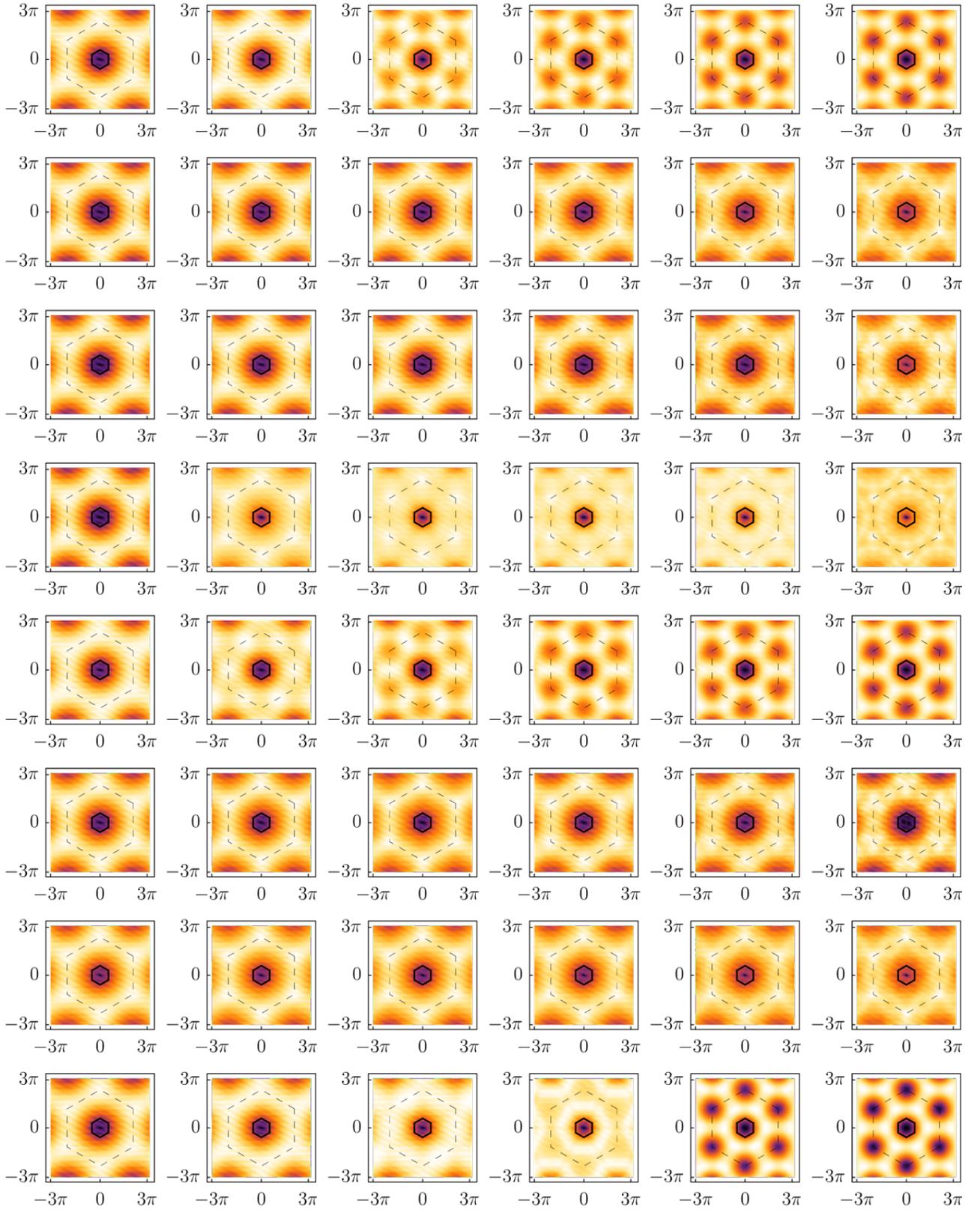

FIG. S4. Same as in Fig. S1 but for the $\mathbb{Z}_2$ *Ansätze* in the $\eta_{C_6} = -\eta = +1$ class.

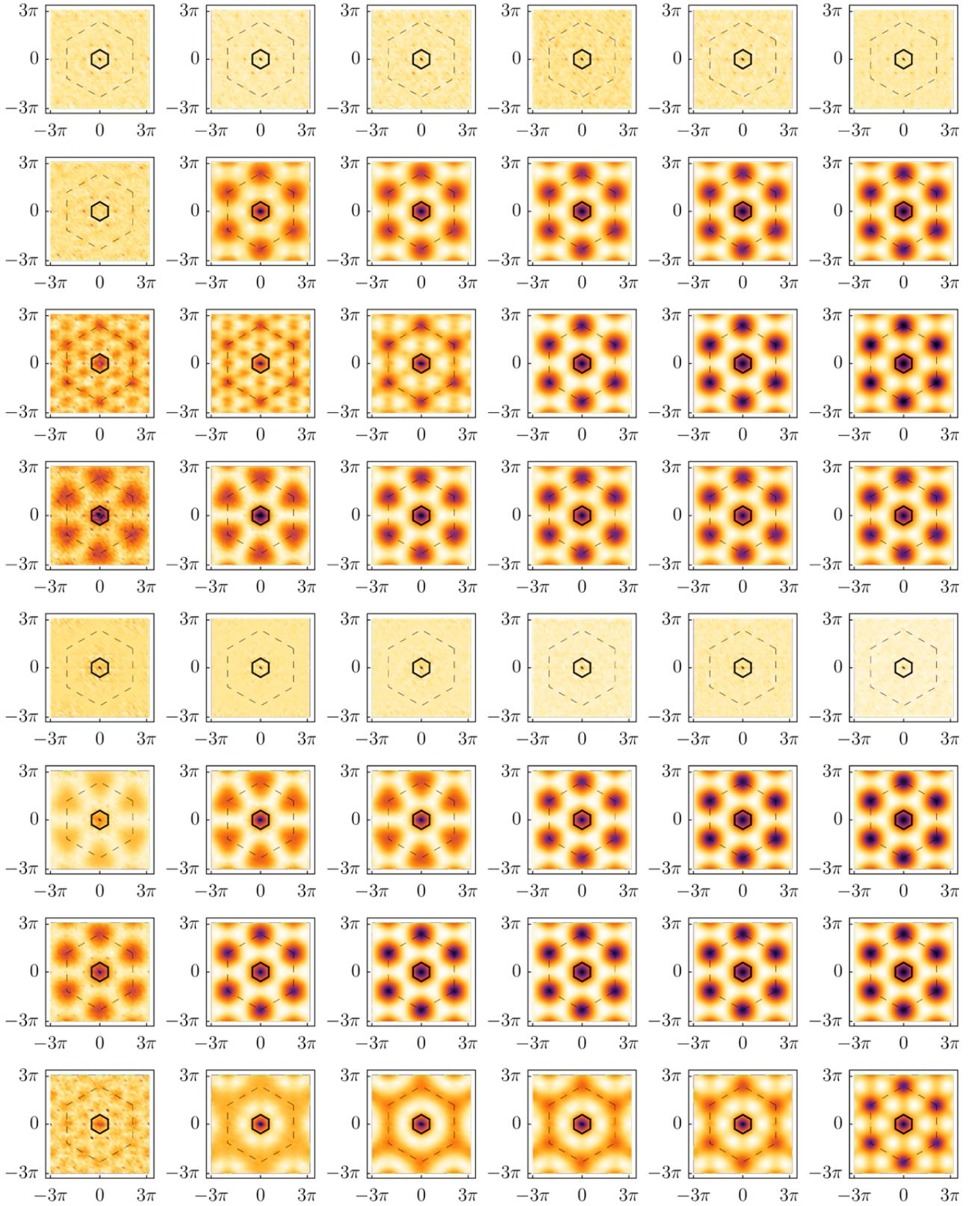

FIG. S5. Static structure factors for the $\mathbb{Z}_2$ *Ansätze* listed in rows 9–12 of Table I, with $\eta_{C_6} = \eta = +1$ (first four rows of the figure) and $\eta_{C_6} = -\eta = -1$ (last four rows here). All other parameters and conventions are chosen to be the same as in Fig. S1.

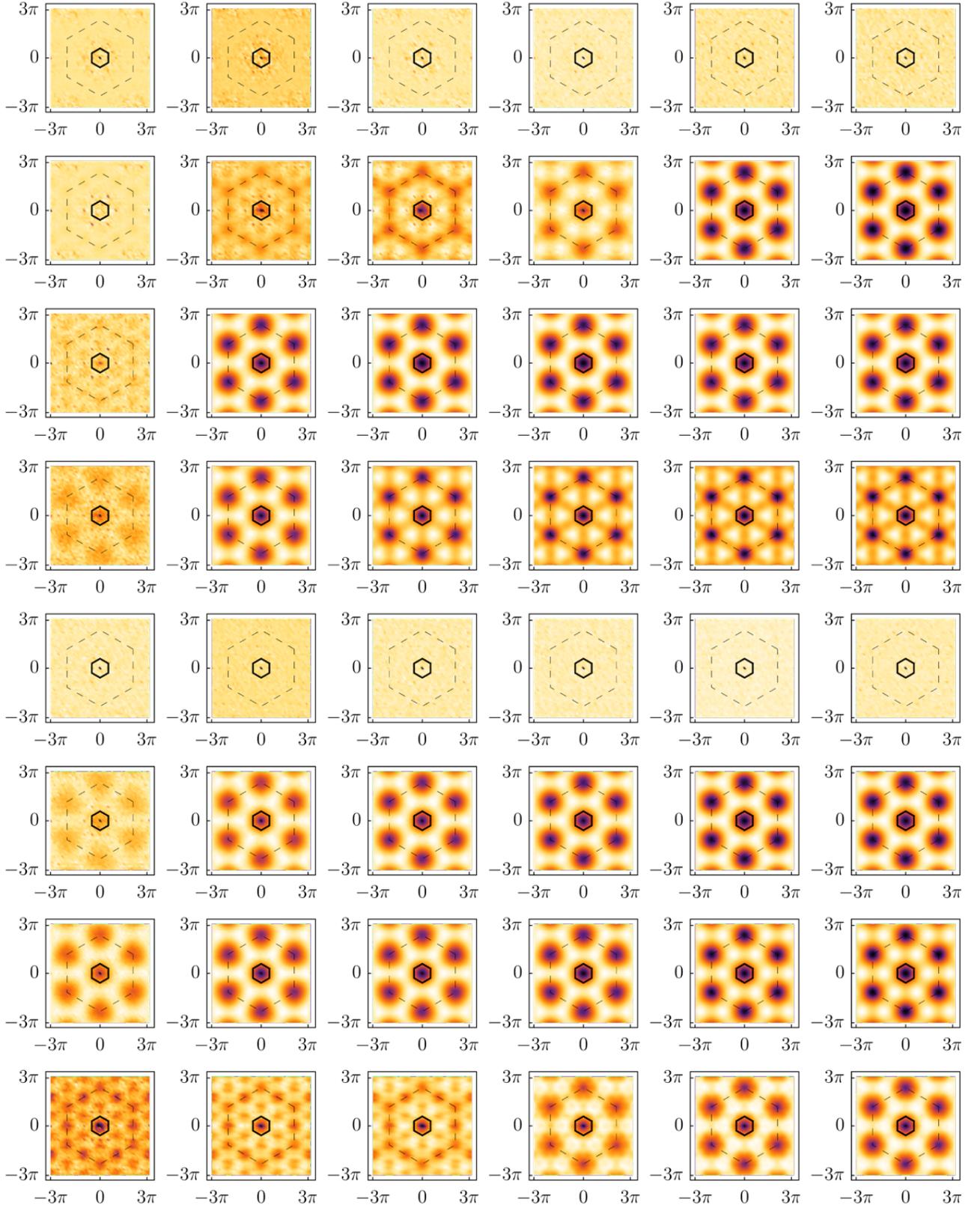

FIG. S6. Same as in Fig. S5, but for the $\eta_{C_6} = \eta = -1$ (first four rows) and $\eta_{C_6} = -\eta = +1$ (last four rows) classes.

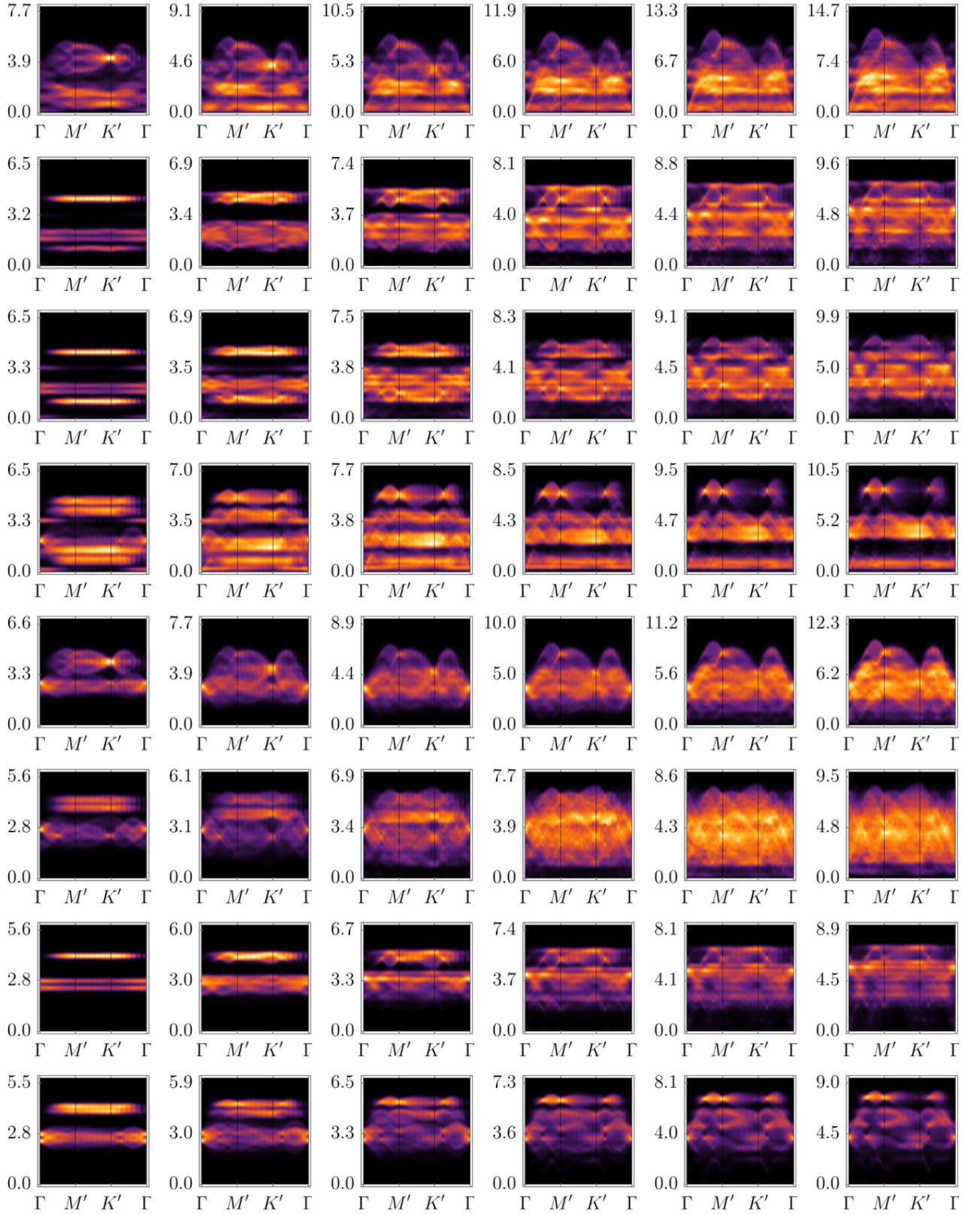

FIG. S7. Dynamical structure factors, plotted along a high-symmetry path in the extended Brillouin zone, for the $\mathbb{Z}_2$ *Ansätze* listed in the first eight rows of Table I, with $\eta_{C_6} = \eta = +1$. All parameters and conventions are chosen to be the same as in Fig. S1.

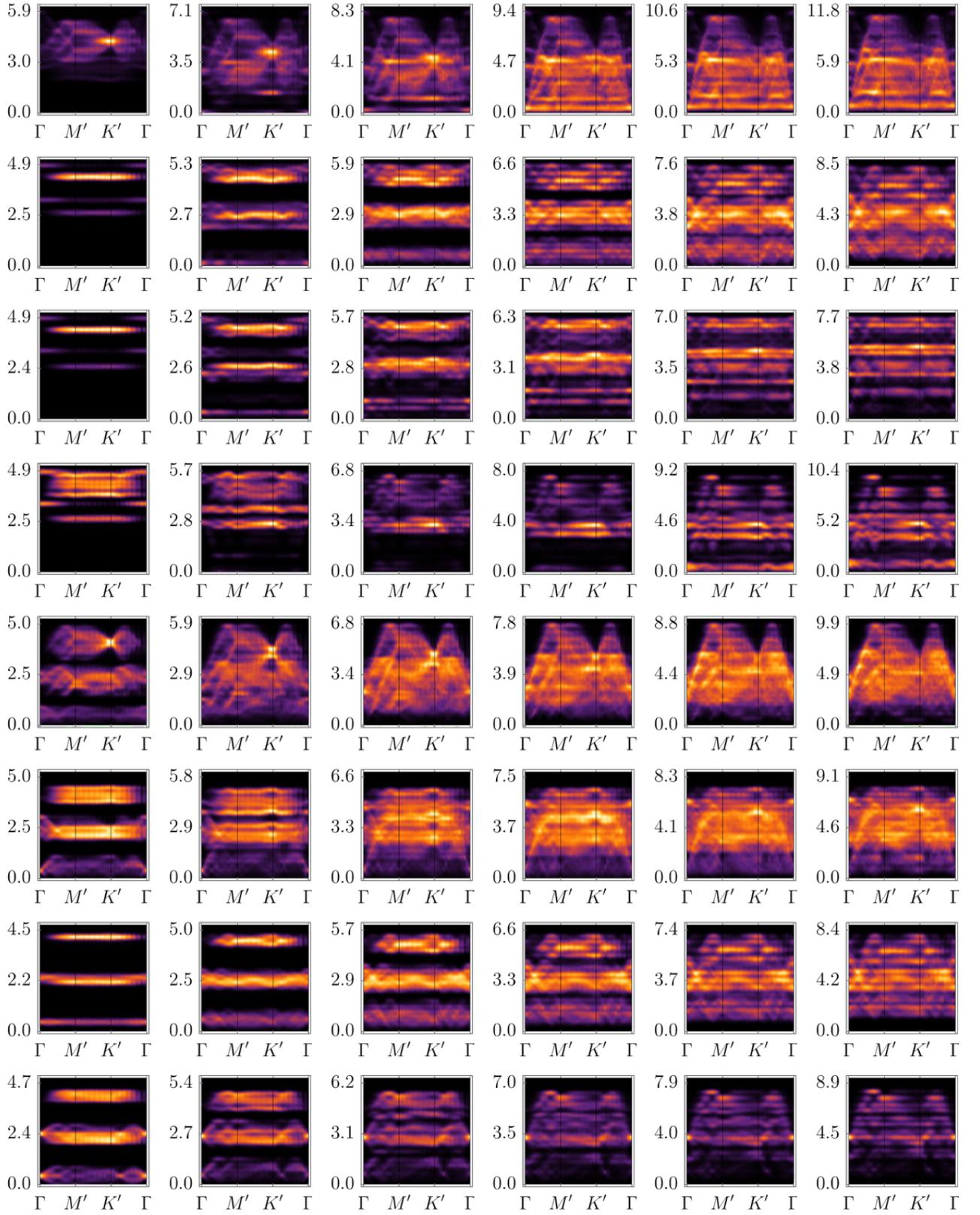

FIG. S8. Same as in Fig. S7 but for the $\mathbb{Z}_2$ *Ansätze* in the $\eta_{C_6} = \eta = -1$ class.

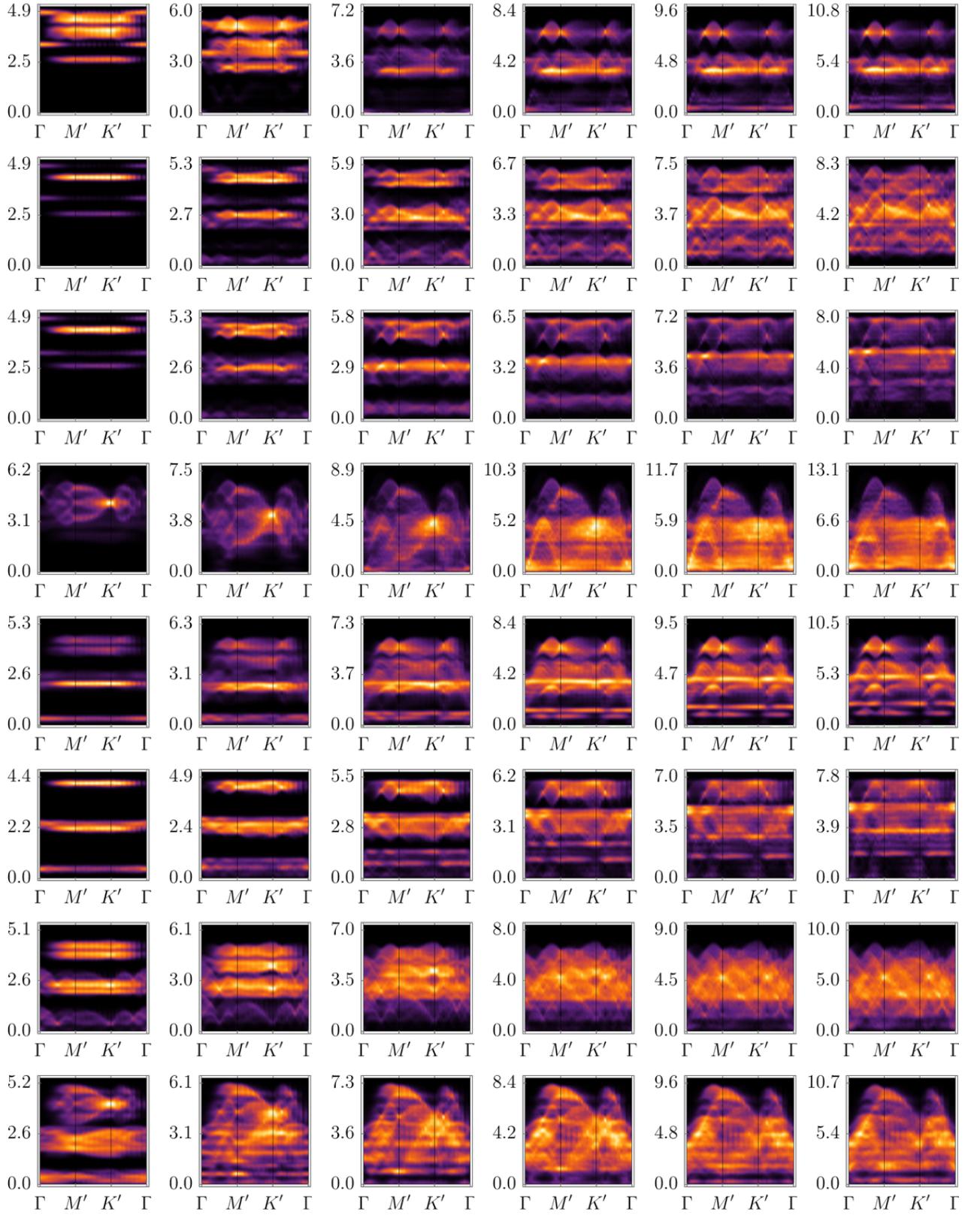

FIG. S9. Same as in Fig. S7 but for the $\mathbb{Z}_2$ *Ansätze* in the $\eta_{C_6} = -\eta = -1$ class.

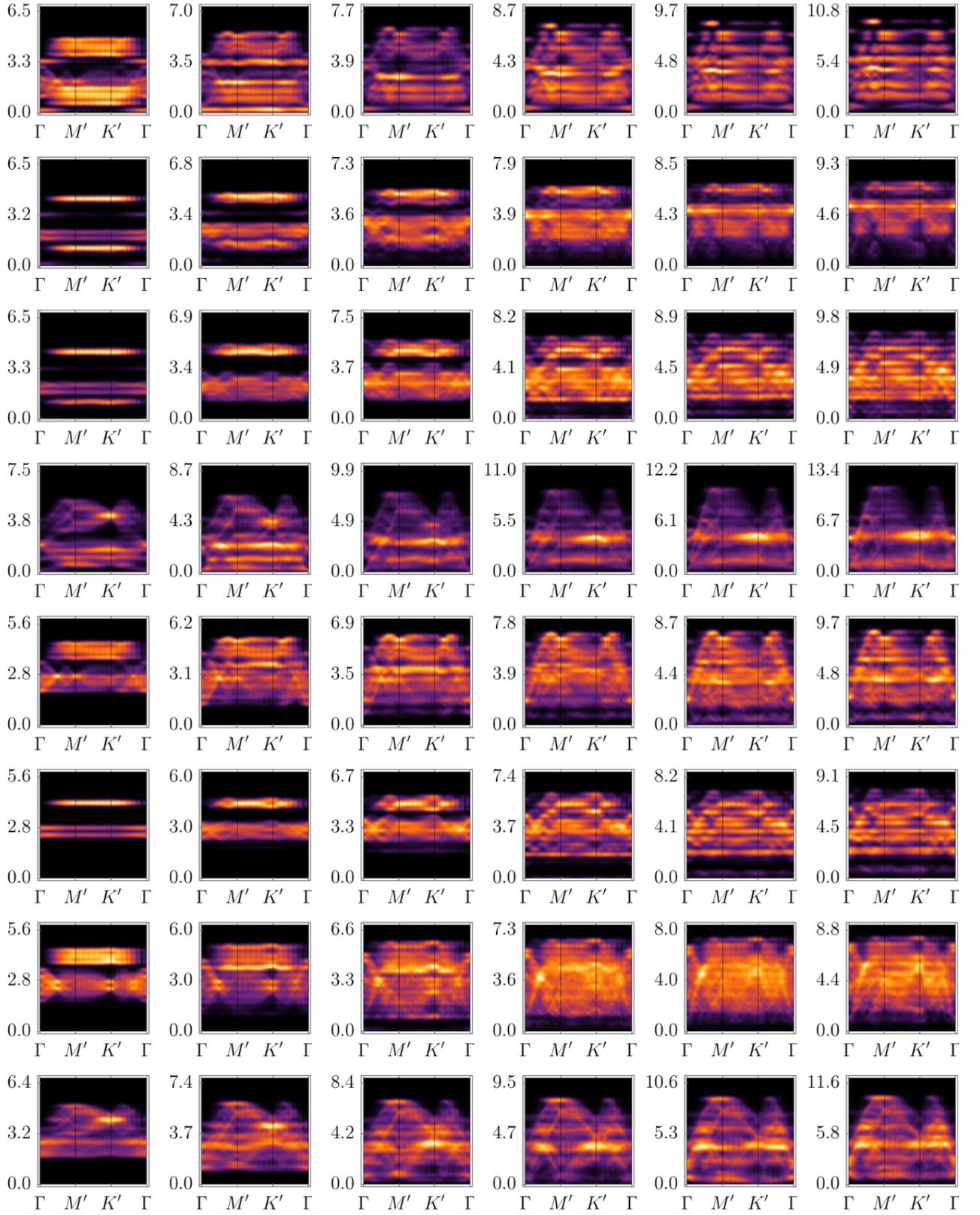

FIG. S10. Same as in Fig. S7 but for the $\mathbb{Z}_2$ *Ansätze* in the $\eta_{C_6} = -\eta = 1$ class.

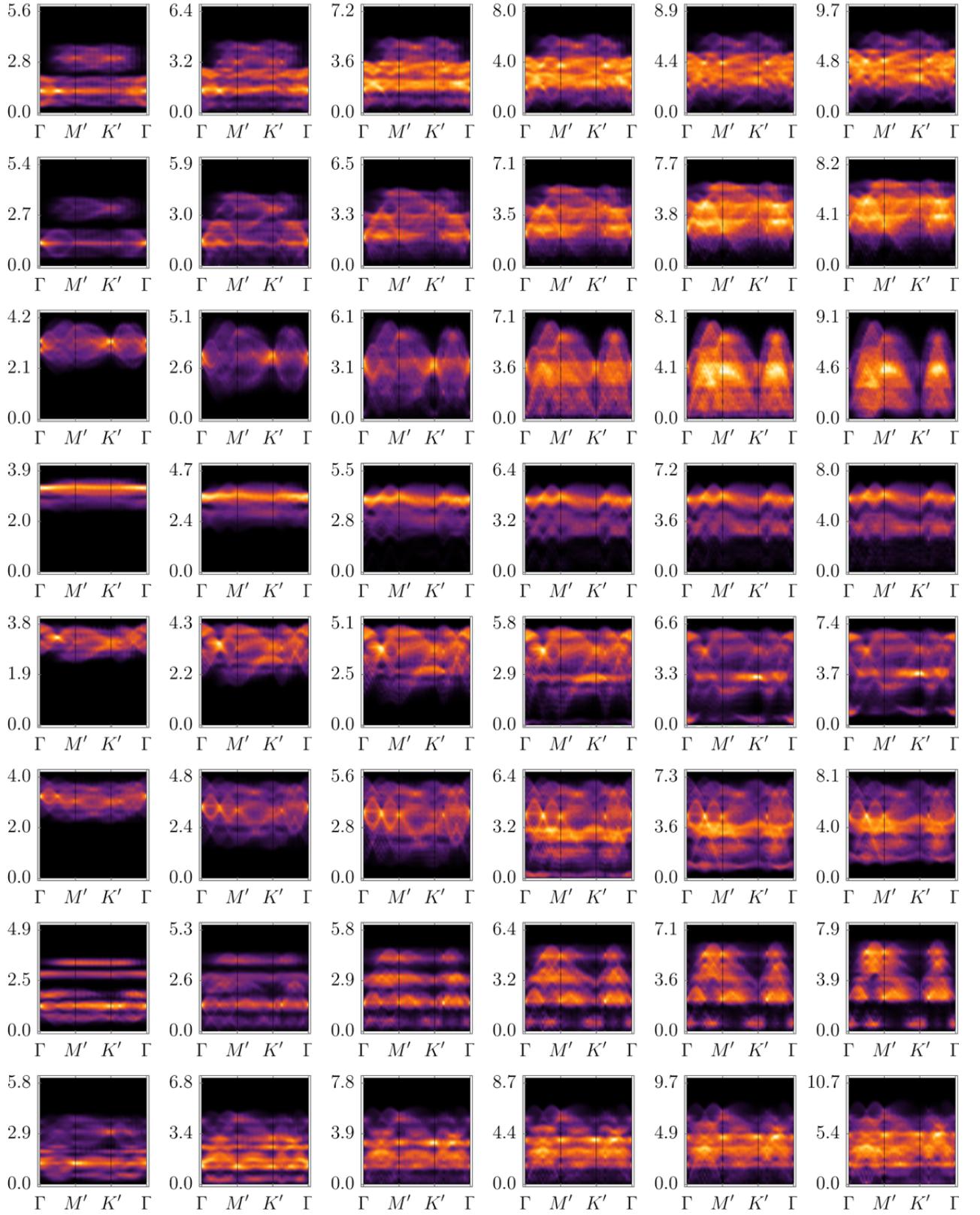

FIG. S11. Dynamical structure factors for the $\mathbb{Z}_2$ *Ansätze* listed in rows 9–12 of Table I, with $\eta_{C_6} = \eta = +1$ (first four rows of the figure) and $\eta_{C_6} = -\eta = -1$ (last four rows here). All other parameters and conventions are taken to be the same as in Fig. S7.

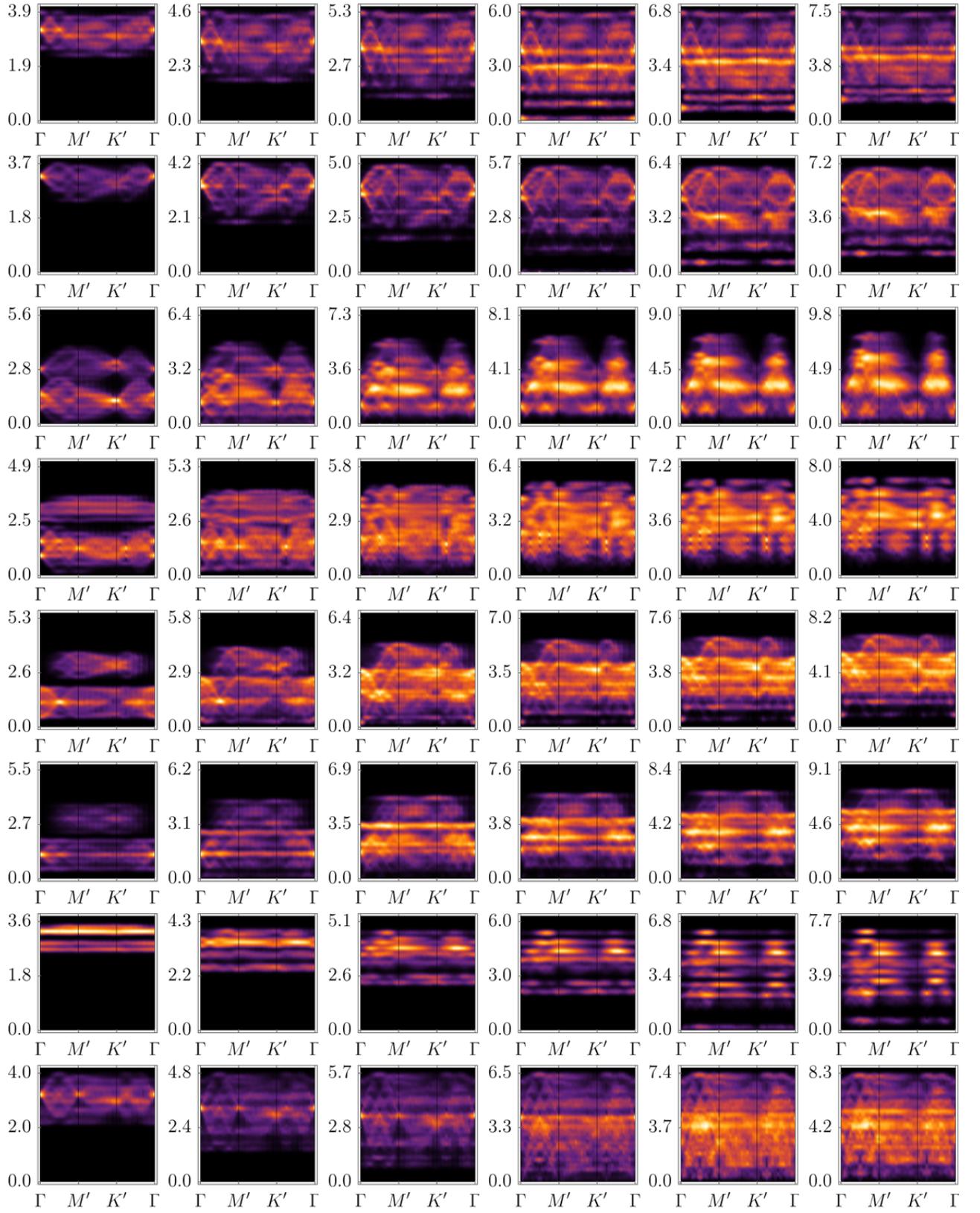

FIG. S12. Same as in Fig. S11 but for the $\eta_{C_6} = \eta = -1$ (first four rows) and $\eta_{C_6} = -\eta = +1$ (last four rows) classes.



\end{document}